\documentclass[12pt]{article}
\usepackage[T1]{fontenc}
\usepackage{tgpagella}
\usepackage[onehalfspacing]{setspace}
\usepackage{parskip}
\usepackage[utf8]{inputenc}
\usepackage{amsmath, amsfonts}
\usepackage{amssymb}
\usepackage{booktabs}
\usepackage{mathtools}

\usepackage{placeins}
\usepackage{graphicx}
\usepackage{xltabular}
\usepackage{longtable}
\usepackage{tabularray }
\usepackage{float}
\usepackage[export]{adjustbox}
\usepackage{caption}
\usepackage{subcaption}
\usepackage[bottom]{footmisc}
\usepackage[margin=1 in]{geometry}
\usepackage{indentfirst}
\usepackage{graphicx} 
\usepackage[table]{xcolor}
\usepackage{lscape}
\usepackage{multirow} 
\usepackage{hyperref}
\usepackage{geometry}
\usepackage{array}
\usepackage{ragged2e}
\usepackage[T1]{fontenc}
\usepackage{tgpagella}
\usepackage[onehalfspacing]{setspace}
\usepackage[utf8]{inputenc}
\usepackage[margin=1in]{geometry}
\usepackage{hyperref}
\hypersetup{colorlinks,linkcolor={blue},citecolor={blue},urlcolor={red}}
\usepackage{natbib}
\numberwithin{figure}{section}
\title{Optimizing Portfolios with Pakistan-Exposed ETFs: Risk and Performance Insight}

\author{
    Ali Jaffri$^1$, Abootaleb Shirvani$^2$, Ayush Jha$^1$, Svetlozar T. Rachev$^4$, Frank J. Fabozzi$^5$ \\
    \\
    {\small $^1$Department of Economics, Texas Tech University} \\
    {\small $^2$Department of Mathematical Sciences, Kean University} \\
    {\small $^4$Department of Mathematical Finance, Texas Tech University} \\
    {\small $^5$Carey Business School, Johns Hopkins University}
}

\date{}
\usepackage[english]{babel}
\begin{document}
\maketitle
\sloppy
\begin{abstract}
This study examines the investment landscape of Pakistan as an emerging and frontier market, focusing on implications for international investors, particularly those in the United States, through exchange-traded funds (ETFs) with exposure to Pakistan. The analysis encompasses 30 ETFs with varying degrees of exposure to Pakistan, covering the period from January 1, 2016, to February 2024. This research highlights the potential benefits and risks associated with investing in these ETFs, emphasizing the importance of thorough risk assessments and portfolio performance comparisons.
By providing descriptive statistics and performance metrics based on historical optimization, this paper aims to equip investors with the necessary insights to make informed decisions when optimizing their portfolios with Pakistan-exposed ETFs. The second part of the paper introduces and assesses dynamic optimization methodologies. This section is designed to explore the adaptability and performance metrics of dynamic optimization techniques in comparison with conventional historical optimization methods. By integrating dynamic optimization into the investigation, this research aims to offer insights into the efficacy of these contrasting methodologies in the context of Pakistan-exposed ETFs.
The findings underscore the significance of Pakistan's market dynamics within the broader context of emerging markets, offering a pathway for diversification and potential growth in investment strategies.
\end{abstract}

\newpage
\section{Introduction}
Pakistan has been categorized in the realm of emerging and frontier markets. The analysis of exchange-traded funds (ETFs) with exposure to Pakistan offers a unique perspective on emerging and frontier market investments in terms of portfolio diversification for international investors, given the changing macroeconomic environment and market dynamics. International investors can take on exposure to Pakistan by using brokerage accounts to purchase scrips directly from the stock market, or they can diversify their portfolios by investing in emerging-market ETFs with exposure to Pakistan. For instance, the Global X MSCI Pakistan ETF (PAK) that was created and managed by Global X Management Company LLC was an ETF that tracked the diverse range of Pakistani firms across sectors like the financial, energy, telecommunication and consumer goods sectors and provided exposure to Pakistan's stock market. This was indeed the first US-listed ETF specifically focused on the Pakistani stock market. 
\onehalfspacing

Emerging markets are ahead of frontier markets in terms of economic growth, market size, liquidity and  having a robust regulatory environment. For instance, the countries in BRICS (Brazil, Russia, India, China and South Africa) are classified as emerging markets. In contrast, countries like Bangladesh, Sri Lanka, Qatar, Oman, etc., are classified in the frontier markets category. Although there is the potential for substantial growth in these markets for investors, frontier markets are considered to be in the high-risk category due to their less developed financial markets and higher volatility.

Pakistan's market status was upgraded from frontier to emerging back in November 2016 by MSCI, but it was downgraded to frontier status in November 2021 due to a reduction in the market size and the illiquidity of the stock market. PAK was launched on April 23, 2015; however, it was de-listed on February 16, 2024. The de-listing has been attributed to various factors, including political and economic uncertainty, low trading volumes, limited market depth, the low liquidity of the assets under management and varying investors' impressions with respect to Pakistan's market classification wavering between the emerging and frontier categories. The key indicators of Pakistan's stock market for the last five years are given in Table 1 and the returns of different asset classes in 2023 are given in Table 2.


\begin{table}[H]
\centering
\begin{adjustbox}{width=1.1\textwidth} 
\begin{tabular}{lccccc}
\rowcolor[HTML]{A2A2B8} 
\multicolumn{1}{c}{\cellcolor[HTML]{A2A2B8}\textbf{Variable}} & \textbf{2020} & \textbf{2021} & \textbf{2022} & \textbf{2023} & \textbf{2024} \\
\hline
\textbf{Total No. of Listed Companies}                       & 531       & 533       & 531       & 524       & 524        \\
\textbf{Total Listed Capital (Rs. in Millions)}               & 1,421,094 & 1,485,103 & 1,552,728 & 1,665,477 & 1,694,457  \\
\textbf{Total Market Capitalization (Rs. in Millions)}        & 8,035,364 & 7,684,637 & 6,500,828 & 9,062,903 & 10,169,955 \\
\textbf{KSE-100™ Index}                                      & 43,755    & 44,596    & 42,420    & 60,451    & 75,878     \\
\textbf{KSE-30™ Index}                                       & 18,180    & 17,502    & 14,836    & 20,777    & 24,343     \\
\textbf{KMI-30 Index}                                        & 71,168    & 71,687    & 68,278    & 104,729   & 125,780    \\
\textbf{KSE All Share Index}                                 & 30,780    & 30,727    & 27,533    & 41,916    & 48,828     \\
\textbf{PSX-KMI All Shares Index}                            & 21,718    & 22,027    & 19,987    & 30,664    & 34,824     \\
\textbf{New Companies Listed During the Year}                & 3         & 7         & 2         & 1         & 6          \\
\textbf{Listed Capital of New Companies (Rs. in Millions)}    & 14,197    & 16,009    & 2,644     & 3,932     & 79,953     \\
\textbf{New Debt Instruments Listed During the Year}         & 7         & 5         & 0         & 5         & 3          \\
\textbf{Listed Capital of New Debt Instruments (Rs. in Millions)} & 246,967   & 25,100    & 0         & 31,200    & 6,075      \\
\textbf{Average Daily Turnover---Regular Market (Shares in Mn, YTD)} & 330       & 474       & 230       & 323       & 450        \\
\textbf{Average Value of Daily Turnover---Regular Market (Rs in Mn, YTD)} & 12,271 & 16,935 & 6,950 & 10,076 & 16,797 \\
\textbf{Average Daily Turnover---Future Market (Shares in Mn, YTD)}       & 102       & 141       & 94        & 106       & 168        \\
\textbf{Average Value of Daily Turnover---Future Market (Rs. in Mn, YTD)} & 4,740     & 8,315     & 3,574     & 4,388     & 6,764      \\
\hline
\end{tabular}
\end{adjustbox}
\caption{Pakistan Stock Exchange Summary of Key Indicators (2020--2024). Source: Pakistan Stock Exchange}
\label{tab:psx-summary}
\end{table}

\begin{table}[H]
\centering
\begin{adjustbox}{width=0.75\textwidth}
\begin{tabular}{lll}
\rowcolor[HTML]{A2A2B8} 
\textbf{Category} & \textbf{Explanation} & \textbf{PKR Return} \\
\hline
KSE-100 & Total Return with Dividend & 53\% \\
Naya Pakistan US\$ Certificate & Including 6.5\% Return & 33\% \\
Commercial Plots Price Index---Karachi & Source: Zameen.com & 29\% \\
US\$ & Interbank Market Rate & 25\% \\
T-Bill & Reinvest after 3 Months & 23\% \\
Gold & Source: Karachi Saraf & 18\% \\
House Price Index---Karachi & Source: Zameen.com & 18\% \\
Bank Saving Deposit & Avg. Bank Rate from SBP & 17\% \\
PIBs (3-Year Bond) & With Coupon & 13\% \\
Special Saving Certificate (SSC) & First-Year Return & 13\% \\
Naya Pakistan PKR Certificate & Investment at Beginning of 2023 & 11\% \\
Residential Plots Price Index---Karachi & Source: Zameen.com & 6\% \\
\hline
\end{tabular}
\end{adjustbox}
\caption{Pakistan Asset Returns in 2023. Source: Topline Securities}
\label{tab:pakistan-asset-returns}
\end{table}
According to \cite{Woetzel2018} and \cite{oecd2019equity}, Asian emerging markets have experienced some of the most robust economic growth rates and outstanding returns in the past, presenting Asia as the world's leading emerging market region. Among these markets, Pakistan has become more prominent. Bloomberg ranked the Pakistan Stock Exchange (PSX) in the top 10 best-performing stock markets in the world for three straight years from 2012 to 2014. Stock markets previously considered as outcasts in the emerging markets world have been among the world's best-performing stock markets during 2024, and Pakistan is one of them, as the market has risen 30\% since the inception of 2024, leaving behind the markets of Taiwan and India \citep{Jilani2024}.

Despite various structural problems, Pakistan's market can still be a good potential avenue for investment for US investors looking for portfolio diversification. Pakistan's economic sector, driven by sectors like textiles, agriculture and the growing IT industry, provides unique exposure. One of the most important aspects of this market is its relatively low correlation with the US and European markets. \cite{Berger2011} studied frontier market equities, including Pakistan, and found that these markets have a low correlation with the world market; hence, they provide diversification opportunities. Using a wavelet-based value-at-risk method, \cite{Mensi2017} found that including a BRIC or South Asian country, especially Pakistan and Sri Lanka, in a portfolio of developed stock markets reduces the resulting portfolio value at risk. \cite{Ngene2018} studied the shock and volatility interactions between the stock markets of  24 frontier markets and the US, and they found that the conditional correlation between the US and each individual frontier market is negative, which can be translated into diversification benefits for US investors. Using the MSCI daily returns data of developed and emerging markets for the period from 2005 to 2018, \cite{Joyo2019} analyzed the correlation between Pakistan and its major trading partners (China, Indonesia, the UK and the US) and concluded that stock markets were strongly correlated during the Great Financial Crisis (GFC), although this integration decreased substantially post-2008.

Studies in the area of portfolio optimization and diversification stress the incorporation of assets with low correlation to reduce the overall risk of the portfolio, presenting Pakistan as an important contender for international investors. In addition to that, Pakistan's young demographic base  and blue economy potential, along with structural reforms, offer more optimistic potential for growth. According to \cite{pwc2017}, over the next three decades, Pakistan will be among the countries with the largest movement in growth, and the forecast predicts that Pakistan could move from $24^{th}$ to $16^{th}$ on the list of top economies around the world by 2050. 

\onehalfspacing

Taking into account the abovementioned factors, the analytics of ETFs with exposure to Pakistan will present noticeable insights into managing risk while finding opportunities in terms of return. For investors particularly seeking to take advantage of geographical and economic diversification, understanding the risk and return dynamics of ETFs can offer substantial benefits. Thus, Pakistan stands as a valuable option for US investors looking to broaden their exposure in emerging and frontier markets. In this paper, we  will study the advantages and disadvantages for US investors investing in ETFs with Pakistan exposure, conducting thorough risk assessments, portfolio analysis and portfolio performance comparisons.


\section{Descriptive Statistics}
\subsection{Data}
Our analyses use different sets of data from Yahoo Finance and FRED. We have selected 30 ETFs with exposure to Pakistan. Table 3 shows the details of each individual ETF. Daily price data on each ETF were obtained from Yahoo Finance, covering the time period from 1/1/2016 to 12/18/2020. Data for the S\&P 500 and the Dow Jones Industrial Average (DJIA) were also extracted from Yahoo Finance. The data for the 3-month treasury yield as a proxy for the risk-free rate were extracted from FRED.

\subsection{ETF Description}



\begin{table}[H]
\resizebox{\textwidth}{!}{%
\begin{tabular}{lllcl}
\rowcolor[HTML]{9B9B9B} 
\multicolumn{1}{c}{\cellcolor[HTML]{9B9B9B}\textbf{Ticker}} &
  \multicolumn{1}{c}{\cellcolor[HTML]{9B9B9B}\textbf{ETF Name}} &
  \textbf{ETF Category} &
  \textbf{Inception Date} &
  \multicolumn{1}{c}{\cellcolor[HTML]{9B9B9B}\textbf{Market Cap (\$bn)}} \\
VWO  & Vanguard FTSE Emerging Markets ETF                      & Emerging Markets Equities         & 4-Mar-05  & 82.95 \\
IEMG & iShares Core MSCI Emerging Markets ETF                  & Emerging Markets Equities         & 18-Oct-12 & 82.40 \\
VXUS & Vanguard Total International Stock ETF                  & Foreign Large-Cap Equities        & 26-Jan-11 & 77.25 \\
IWM  & iShares Russell 2000 ETF                                & Small-Cap Blend Equities          & 22-May-00 & 68.34 \\
VT   & Vanguard Total World Stock ETF                          & Large-Cap Growth Equities         & 24-Jun-08 & 40.35 \\
VEU  & Vanguard FTSE All-World ex-US ETF                       & Foreign Large-Cap Equities        & 2-Mar-07  & 39.75 \\
IXUS & iShares Core MSCI Total International Stock ETF         & Foreign Large-Cap Equities        & 18-Oct-12 & 38.07 \\
EEM  & iShares MSCI Emerging Markets ETF                       & Emerging Markets Equities         & 7-Apr-03  & 18.12 \\
IWN  & iShares Russell 2000 Value ETF                          & Small-Cap Blend Equities          & 24-Jul-00 & 12.38 \\
IWO  & iShares Russell 2000 Growth ETF                         & Small-Cap Growth Equities         & 24-Jul-00 & 12.02 \\
VSS  & Vanguard FTSE All-World ex-US Small-Cap ETF             & Foreign Small- \& Mid-Cap Equities & 2-Apr-09  & 8.75  \\
ACWX & iShares MSCI ACWI ex U.S. ETF                           & Foreign Large-Cap Equities        & 26-Mar-08 & 4.59  \\
EEMV & iShares MSCI Emerging Markets Min Vol Factor ETF        & Asia Pacific Equities             & 18-Oct-11 & 4.24  \\
AAXJ & iShares MSCI All Country Asia ex Japan ETF              & Asia Pacific Equities             & 13-Aug-08 & 2.59  \\
IWC  & iShares Micro-Cap ETF                                   & Small-Cap Blend Equities          & 12-Aug-05 & 0.91  \\
SPGM & SPDR Portfolio MSCI Global Stock Market ETF             & Global Equities                   & 27-Feb-12 & 0.88  \\
EWX  & SPDR S\&P Emerging Markets Small Cap ETF                & Emerging Markets Equities         & 27-May-08 & 0.75  \\
EEMA & iShares MSCI Emerging Markets Asia ETF                  & Asia Pacific Equities             & 8-Feb-12  & 0.48  \\
GMF  & SPDR S\&P Emerging Asia Pacific ETF                     & Asia Pacific Equities             & 20-Mar-07 & 0.37  \\
EEMS & iShares MSCI Emerging Markets Small-Cap ETF             & Foreign Small- \& Mid-Cap Equities & 16-Aug-11 & 0.36  \\
JPEM & JPMorgan Diversified Return Emerging Markets Equity ETF & Emerging Markets Equities         & 7-Jan-15  & 0.32  \\
TLTE & FlexShares Morningstar Emerging Markets Factor Tilt Index Fund & Foreign Large-Cap Equities & 25-Sep-12 & 0.28 \\
HEEM & iShares Currency Hedged MSCI Emerging Markets ETF       & Emerging Markets Equities         & 29-Jun-15 & 0.17  \\
VEGI & iShares MSCI Global Agriculture Producers ETF           & Commodity Producers Equities      & 31-Jan-12 & 0.10  \\
FILL & iShares MSCI Global Energy Producers ETF                & Energy Equities                   & 12-Sep-11 & 0.09  \\
CUT  & Invesco MSCI Global Timber ETF                          & Materials                         & 19-Nov-07 & 0.05  \\
QEMM & SPDR MSCI Emerging Markets StrategicFactors ETF         & Emerging Markets Equities         & 4-Jun-14  & 0.05  \\
SDEM & Global X MSCI SuperDividend Emerging Markets ETF        & Emerging Markets Equities         & 16-Mar-15 & 0.04  \\
SMCP & Alpha Architect International Quantitative Value ETF    & Small-Cap Blend Equities          & 27-Dec-17 & 0.04  \\
EEMO & Invesco S\&P Emerging Markets Momentum ETF              & Emerging Markets Equities         & 24-Feb-12 & 0.01  \\
\hline
\end{tabular}%
}
\caption{ETF Details with Market Capitalization}
\label{tab:my-table}

\end{table}

The daily returns for each ETF were computed from the price data. To compare the performances of the different ETFs, we computed a cumulative investment price for each ETF, PAK and EQW, assuming a \$100  (long-only) investment in each on 1/1/2016.

\begin{figure}[H]
\captionsetup{justification=centering}
\centering 
\includegraphics[width=1.0\textwidth]{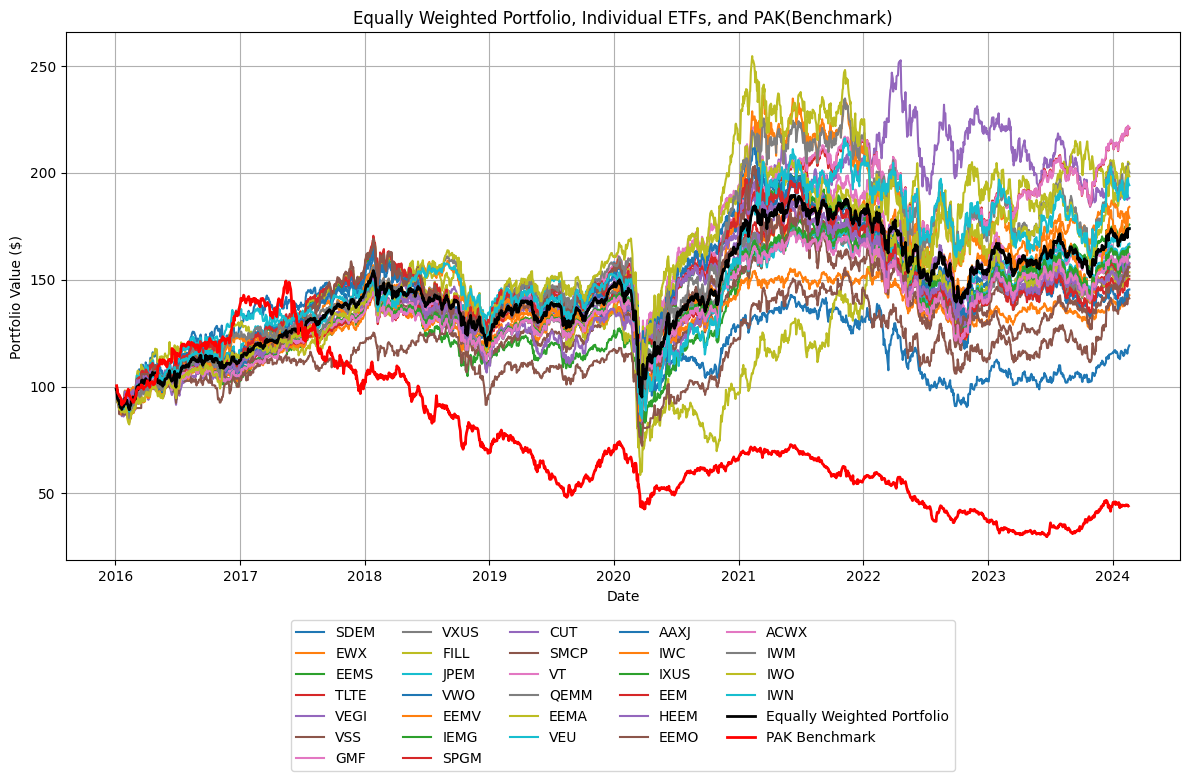}
    \caption[The system.]{Cumulative prices for US-traded ETFs, PAK and EQW. Each time series assumes a \$100 investment on Jan. 1, 2016.}
\label{theSystemFig}
\end{figure}

We make the following observations on the performance of these  ETFs during this time period:

\begin{enumerate}
    \item The PAK benchmark (highlighted in red) has underperformed significantly relative to the equally weighted portfolio and the individual ETFs. While most ETFs maintained or increased their value over time, the PAK benchmark saw a marked decline, suggesting poor performance in either the underlying market or sector compared to the other ETFs.
    \item EQW (black line) shows a more stable trajectory, with moderate growth over time. It is generally less volatile compared to individual ETFs, indicating that diversification among these ETFs helped reduce risk and provided a buffer against the extreme fluctuations seen in some individual assets.
    \item Around the beginning of 2020, there is a noticeable dip across all assets, likely due to the COVID-19 market crash. However, the majority of ETFs and the equally weighted portfolio recovered quickly, showing resilience and growth in subsequent periods, which might indicate a strong recovery across the sectors represented in the portfolio. The clustering of most ETFs towards the top of the chart by 2024 suggests a strong overall market performance, despite the continued underperformance of the PAK benchmark. 
\end{enumerate}

In this paper, we will use EQW as a benchmark.

\section{Historical Portfolio Optimization}

This section analyzes the different asset allocation tools institutional investment managers use to investigate the different risk-return profiles to accommodate various market environments and risk tolerances for the ETF presented in this paper. Given the performance comparison of the equally weighted portfolio (EWP) against PAK, the ETF with the highest exposure to Pakistan's financial market, we conduct a historical analysis based on the method outlined in \citet{reit-portfolio}. To investigate the performance of our portfolio, we construct an EWP and a Markowitz efficient frontier that is robust in conducting a historical analysis.

Given a portfolio with $N$ assets, the weight $w_{i}$ assigned to each asset $i$ in an EWP is

\begin{equation}
    w_{i} = \frac{1}{N},\;\;\;\forall\;\;i = 1, 2, \hdots, N,
\end{equation}

given 

\begin{equation*}
    \sum_{i=1}^{N} w_{i} = 1.
\end{equation*}

If $R_{i}$ is the return of each asset $i$, then the return of the EWP, $R_{p}$, is

\begin{equation}
    R_{p} = \sum_{i=1}^{N}w_{i}R_{i} = \frac{1}{N}\sum_{i=1}^{N}R_{i}.
\end{equation}

Using an EWP as a benchmark to analyze the historical performance of the PAK ETF (an exchange-traded fund with the highest exposure to Pakistan) offers a compelling approach that can be used to evaluate the risk and return profile of a concentrated, country-specific investment. An EWP, by definition (see \citet{markowitz} and \citet{investments}), allocates an equal proportion of investment capital to each included asset, providing a neutral, diversified baseline. This benchmark does not favor any specific sector or country and thus stands as an effective comparison point for the single-country focus of PAK. Evaluating PAK against an EWP allows an assessment of how concentrated exposure to Pakistan’s market measures up against a diversified strategy, especially regarding the risk, return and volatility.

In this section, we will consider the performance of the optimizations on the portfolio of 30 ETFs under a long-only strategy and a basic long-short strategy. Weights for the individual ETFs are determined based on the returns from a rolling window of 1,008 trading days (four trading years). The time window will give us a sample large enough to create a feasible set of values of weights. After constructing complete time series of optimized portfolio weights, we computed performance measures. We are not following historical optimization for the weights of EQW; instead, the weights are computed based on the equal weighting of the prices of the assets in the portfolio on the previous day. 

\subsection{Basic Strategies, Price and Return Performance}
\subsubsection{Long Only }
The performance of the cumulative price of each portfolio from 11/13/2019 through 2/19/2024, assuming a \$100 investment in the portfolio on 11/12/2019, is shown in Figure 3.1. As shown by the plot, tangent portfolios, including the time-varying portfolio (TVP), T95 and T99,  outperform all the others. The minimum variance portfolio (MVP) and C95 strongly track each other. Interestingly, the unoptimized EQW portfolio performs rather well, most noticeably in the post-crisis period from 2021--2024. However, in the long term it underperforms the tangent portfolios while outperforming the global risk-minimizing portfolios.

\begin{figure}[H]
\captionsetup{justification=centering}
\centering
\includegraphics[width=1.0\textwidth]{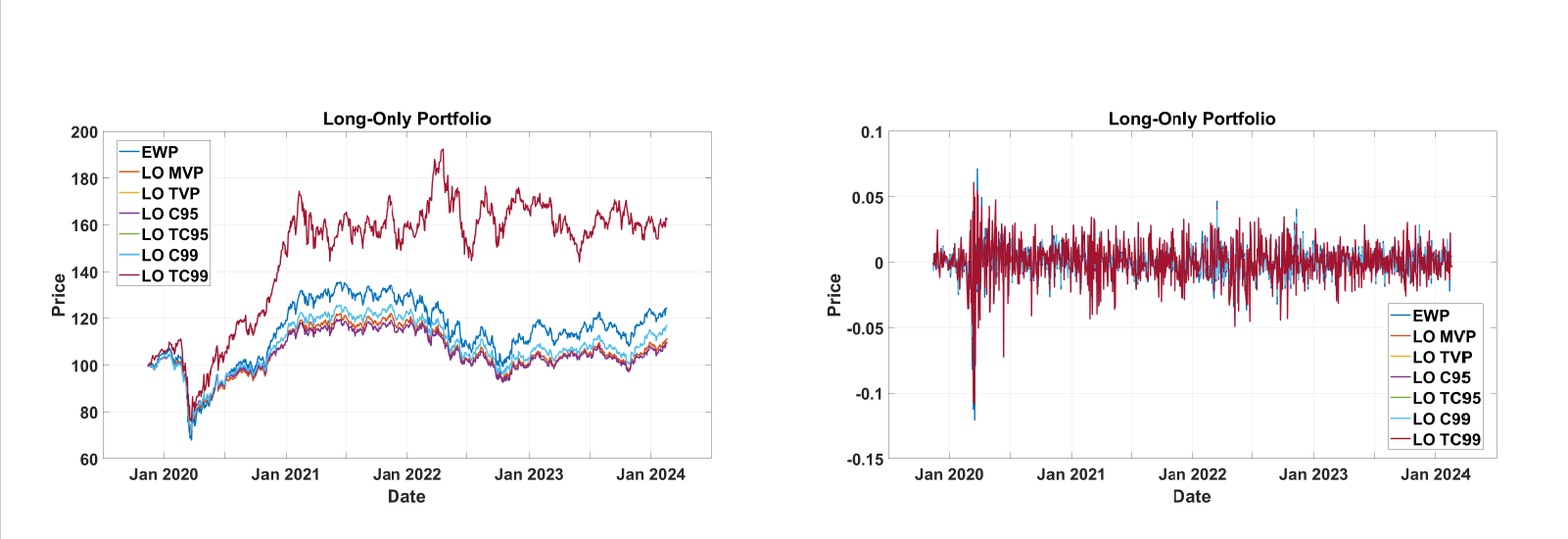}
\caption[The system.]{Comparison of the cumulative price (left) and log-return (right) of the long-only portfolios to those of the benchmark.}
\label{theSystemFig}
\end{figure}

\subsubsection{Long Short}

The evaluation of long-short portfolios constructed using 30 ETFs with an initial investment of \$100 reveals distinct risk-return trade-offs. The long-short (LS) TC99 portfolio significantly outperforms the others, showcasing its strong growth potential. However, this portfolio exhibits substantial volatility, as evidenced in the right panel. Portfolios like LS C95 and LS TC95 offer moderate returns with relatively low volatility, striking a balance between risk and reward. On the other hand, more conservative strategies such as the EWP, LS MVP and LS TVP show stable performance but limited growth, barely exceeding the  baseline. These strategies are well-suited for risk-averse investors prioritizing capital preservation over returns. 
The right panel emphasizes that aggressive portfolios like LS TC99 (long-short tracking constraint 99\%) and LS C95 have highly volatile returns, whereas the EWP and LS MVP deliver stable, consistent returns. Overall, LS TC99 offers the highest returns at the cost of significant volatility, while the EWP and LS MVP prioritize stability for conservative investors.

\begin{figure}[H]
\captionsetup{justification=centering}
\centering
\includegraphics[width=1.0\textwidth]{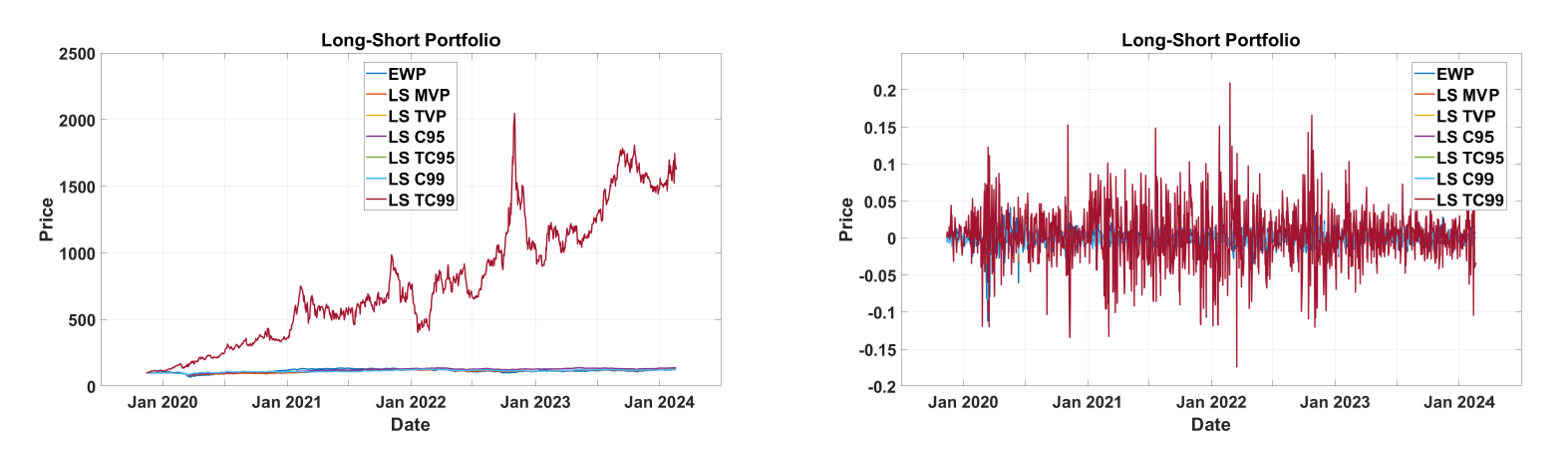}
\caption[The system.]{Comparison of the cumulative price (left) and log-return (right) of the long-short portfolios to those of the benchmark.}
\label{theSystemFig}
\end{figure}

\subsection{Tail Risk Comparison}

In this section, we will compare the tail risk using the Hill estimator. The Hill estimator is often used to estimate the tail index of financial indices. The purpose is to comprehend the probability of extreme losses or gains and investigate the distribution of the tails. For this study, we focus on the following comparisons:

\begin{enumerate}
    \item Equal-weighted portfolio (EWP) vs. Dow Jones Industrial Average (DJIA);
    \item  Global X MSCI Pakistan ETF (PAK)  vs. S\&P 500.
\end{enumerate}

\begin{figure}[H]
\captionsetup{justification=centering}
\centering
\includegraphics[width=1.0\textwidth]{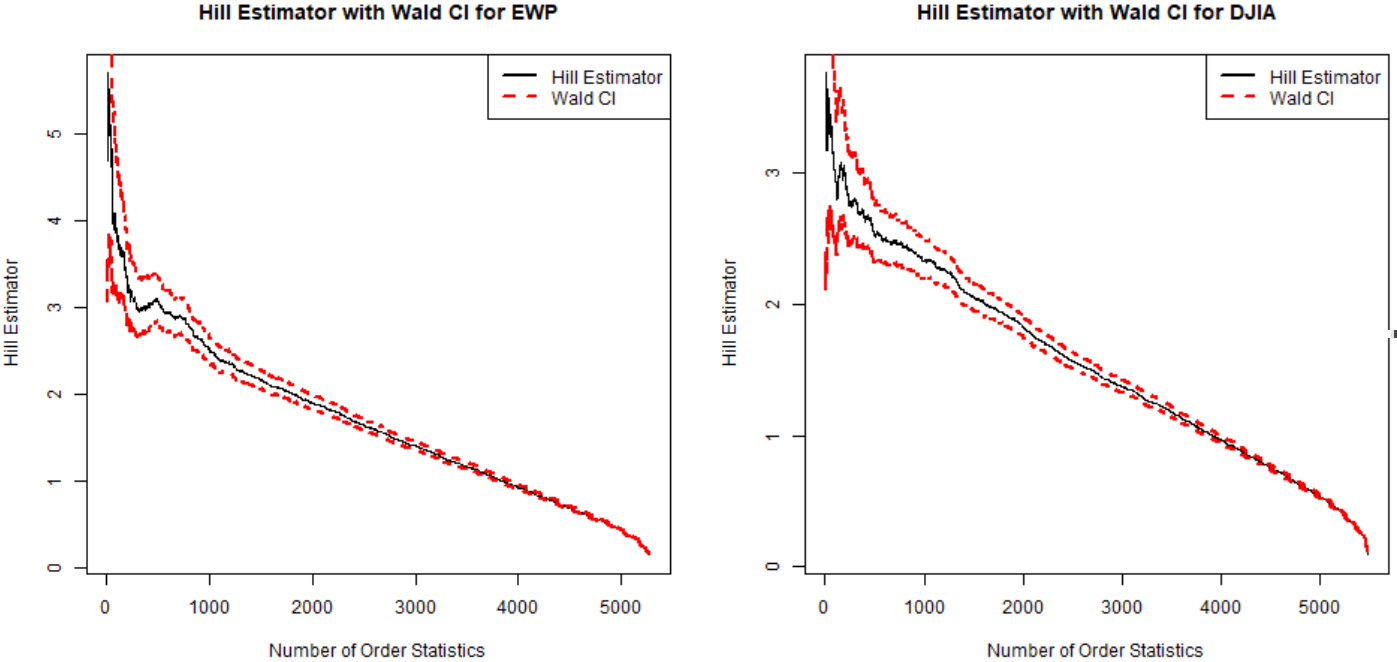}
\caption[The system.]{EWP (left panel) and DJIA (right panel) estimated tail index, along with the Wald confidence interval (CI).}
\label{theSystemFig}
\end{figure}

As we can see from Figure 3.3, the EWP has a higher initial Hill estimator value compared to the DJIA, indicating a potentially heavier tail for the initial order statistics. This could mean that extreme events in the EWP are more pronounced compared to those in the DJIA, possibly due to the equal-weighted nature of the portfolio, which might introduce more variability. As the number of order statistics increases, both the EWP and DJIA show a stabilization in their Hill estimators, converging towards a steady estimate. However, the EWP exhibits more fluctuation early on compared to the DJIA, suggesting that the DJIA may have more stable tail behavior.

Similarly, in Figure 3.4, the Hill estimator for the S\&P 500 is notably lower compared to the market cap portfolio. It starts around a value of 3.5 and then gradually decreases and stabilizes as more order statistics are included. This lower value indicates that the S\&P 500 has a lighter tail compared to the market cap portfolio, implying fewer extreme events in the distribution of returns. The Wald confidence interval for the S\&P 500 is initially wider but narrows more quickly than that of the market cap portfolio. This suggests that the tail estimation for the S\&P 500 becomes more reliable with fewer extreme values, indicating a more stable tail distribution.

\begin{figure}[H]
\captionsetup{justification=centering}
\centering
\includegraphics[width=1.0\textwidth]{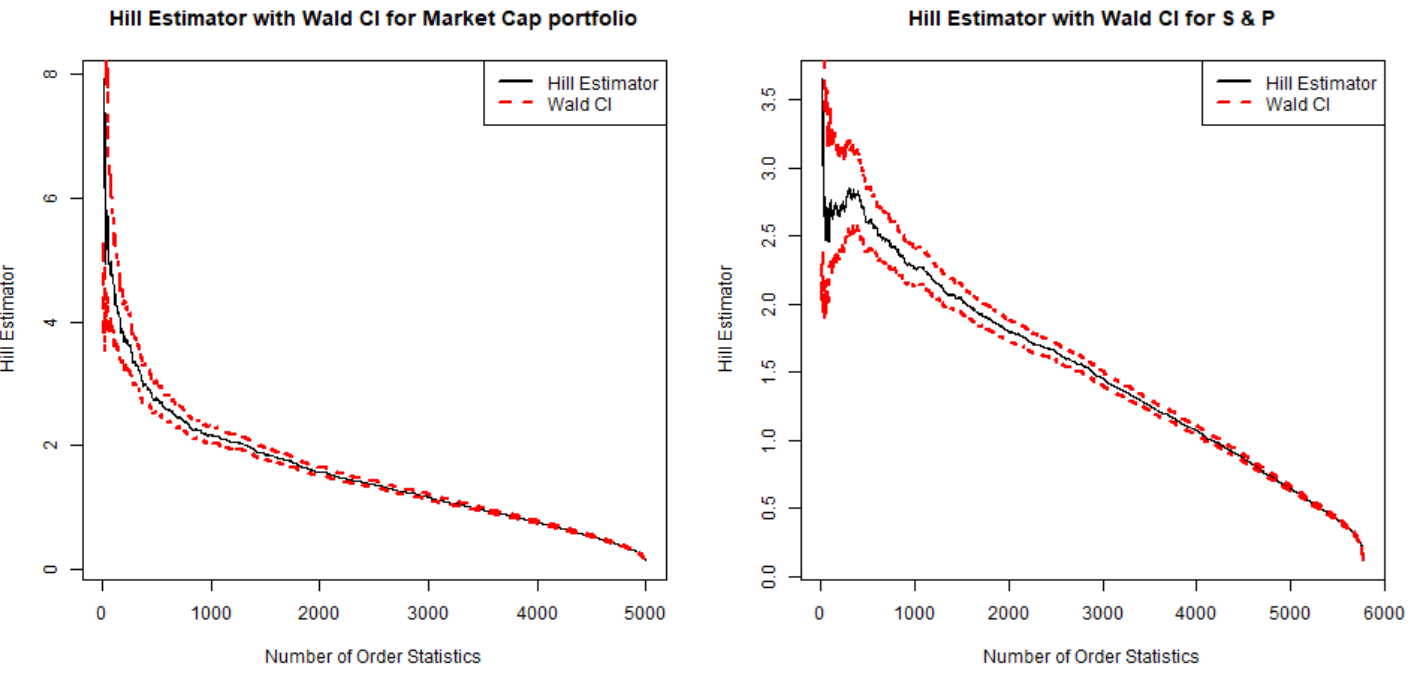}
\caption[The system.]{PAK (left panel) and S\&P 500 (right panel) estimated tail index, along with the Wald confidence interval (CI).}
\label{theSystemFig}
\end{figure}
\newpage
\subsection{Robust Regression}

In this section, we aim to understand the relationship between the returns of specific ETFs and their respective benchmarks. By regressing ETF returns against benchmark returns, we can determine each ETF's sensitivity to the benchmark's movements, gaining insights into the ETF's risk and return profile relative to the broader market or sector it tracks. This analysis is essential for investors and researchers seeking to understand how ETFs perform in various market conditions and how they are correlated with their benchmarks.

The regression model used here can be expressed mathematically as follows:

\begin{equation*}
    Y_{\text{ETF}} = \alpha + \beta \cdot X_{\text{Benchmark}} + \epsilon.
\end{equation*}

\begin{itemize}
    \item \( Y_{\text{ETF}} \) represents the returns of a specific ETF (e.g., VWO, IEMG).
    \item \( X_{\text{Benchmark}} \) denotes the returns of a specific benchmark (either EWP or PAK in this study).
    \item \( \alpha \) (alpha) is the intercept, representing the expected return of the ETF when the benchmark return is zero.
    \item \( \beta \) (beta) is the slope, measuring the sensitivity of the ETF’s returns to the benchmark’s returns.
    \item \( \epsilon \) is the error term, capturing the deviations of actual returns from the fitted line.
\end{itemize}

Using the Huber T norm, we perform a robust regression for each ETF-benchmark pair. This technique minimizes the impact of extreme values while maintaining the core structure of the linear relationship between the ETF and benchmark returns. We calculate the 95\% Wald confidence intervals for the regression estimates. These intervals provide a range of values within which we can be reasonably confident the true values of alpha and beta lie.

\begin{figure}[H]
    \centering
    \begin{minipage}{0.45\textwidth}
        \centering
        \includegraphics[width=\linewidth]{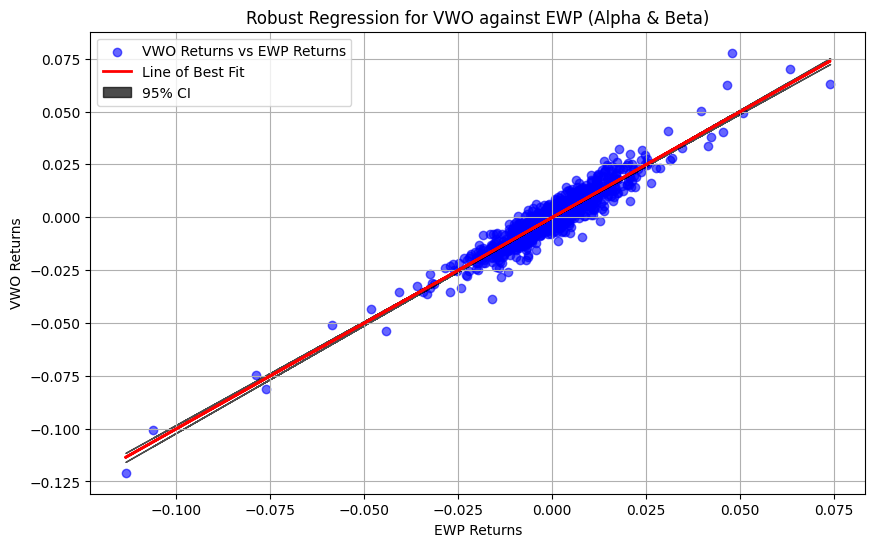}
        \captionsetup{labelformat=empty} 
        \caption{VWO}
    \end{minipage}%
    \hspace{0.05\textwidth}
    \begin{minipage}{0.45\textwidth}
        \centering
        \includegraphics[width=\linewidth]{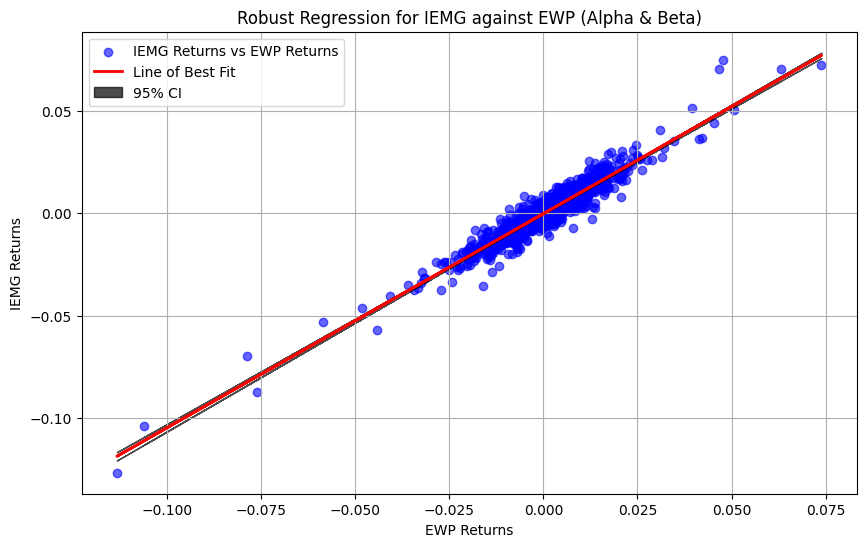}
        \captionsetup{labelformat=empty} 
        \caption{IEMG}
    \end{minipage}
    
    \vspace{0.5cm}
    
    \begin{minipage}{0.45\textwidth}
        \centering
        \includegraphics[width=\linewidth]{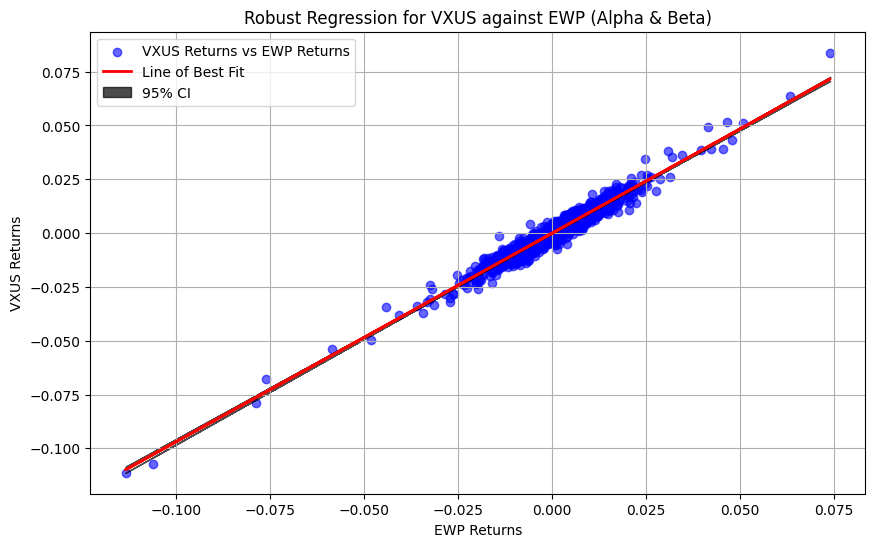}
        \captionsetup{labelformat=empty} 
        \caption{VXUS}
    \end{minipage}%
    \hspace{0.05\textwidth}
    \begin{minipage}{0.45\textwidth}
        \centering
        \includegraphics[width=\linewidth]{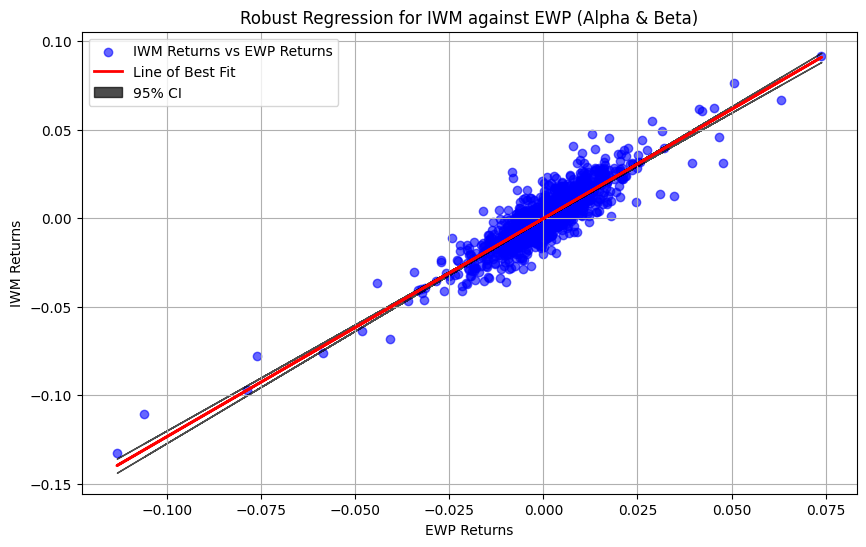}
        \captionsetup{labelformat=empty} 
        \caption{IWM}
    \end{minipage}
    
    \vspace{0.5cm}
    
    \begin{minipage}{0.45\textwidth}
        \centering
        \includegraphics[width=\linewidth]{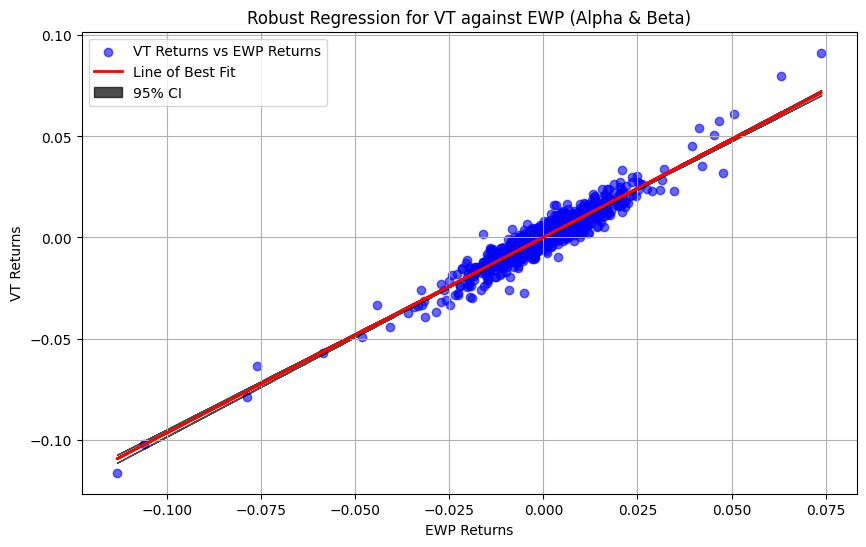}
        \captionsetup{labelformat=empty} 
        \caption{VT}
    \end{minipage}%
    \hspace{0.05\textwidth}
    \begin{minipage}{0.45\textwidth}
        \centering
        \includegraphics[width=\linewidth]{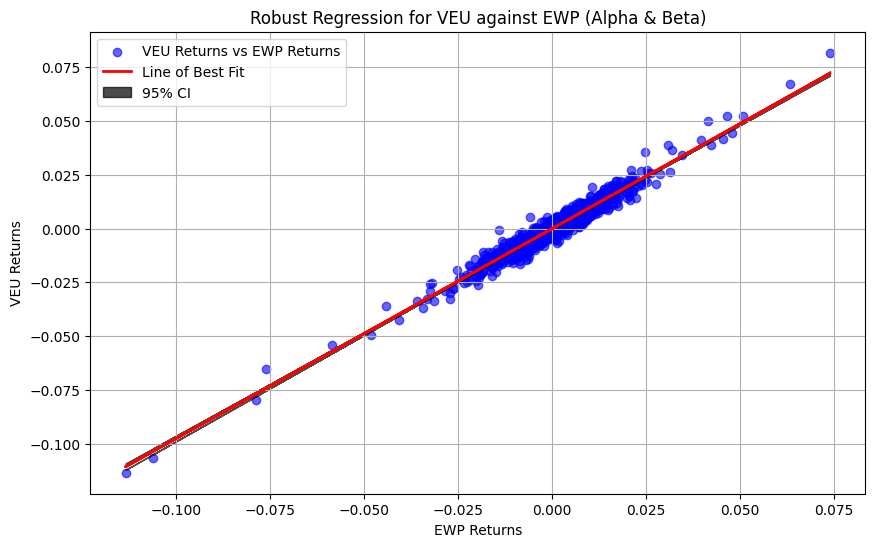}
        \captionsetup{labelformat=empty} 
        \caption{VEU}
    \end{minipage}
    
    \vspace{0.5cm}
    
    \begin{minipage}{0.45\textwidth}
        \centering
        \includegraphics[width=\linewidth]{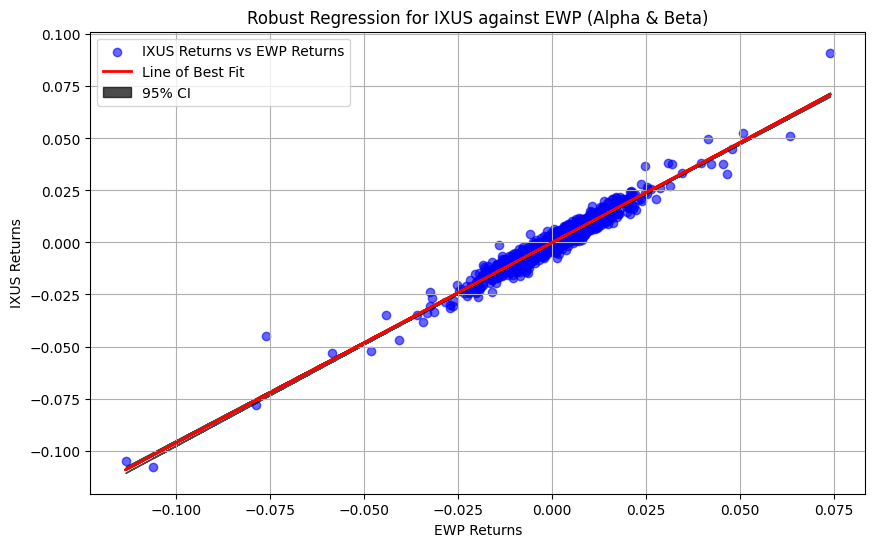}
        \captionsetup{labelformat=empty} 
        \caption{IXUS}
    \end{minipage}%
    \hspace{0.05\textwidth}
    \begin{minipage}{0.45\textwidth}
        \centering
        \includegraphics[width=\linewidth]{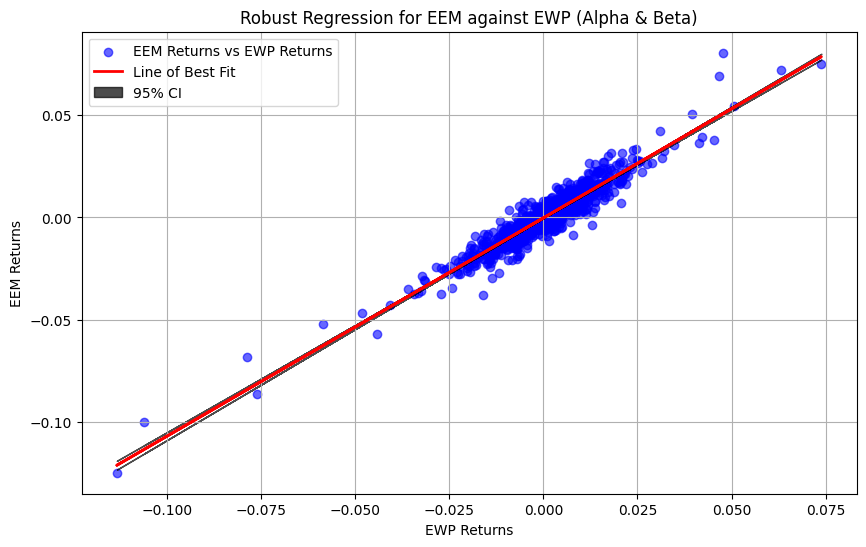}
        \captionsetup{labelformat=empty} 
        \caption{EEM}
    \end{minipage}
\end{figure}

\clearpage

\begin{figure}[H]
    \centering
    \begin{minipage}{0.45\textwidth}
        \centering
        \includegraphics[width=\linewidth]{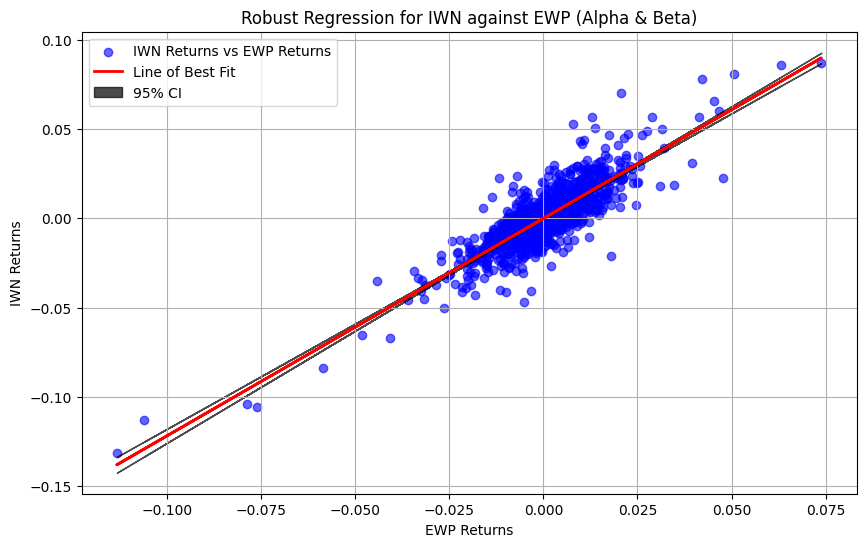}
        \captionsetup{labelformat=empty} 
        \caption{IWN}
    \end{minipage}%
    \hspace{0.05\textwidth}
    \begin{minipage}{0.45\textwidth}
        \centering
        \includegraphics[width=\linewidth]{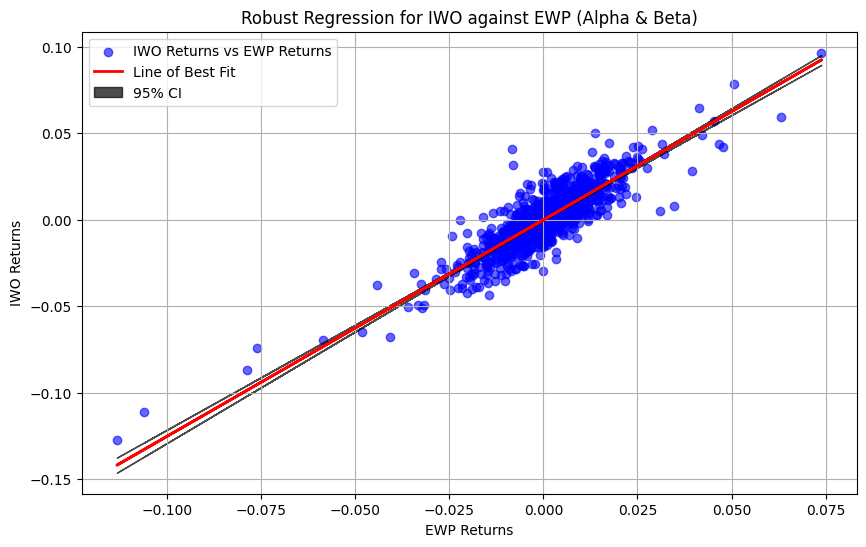}
        \captionsetup{labelformat=empty} 
        \caption{IWO}
    \end{minipage}
\end{figure}

The ETFs show a consistently strong positive relationship with the EQW benchmark. Most ETFs have points tightly clustered around the regression line, with narrow confidence intervals, indicating a stable and predictable relationship. Slopes (beta values) are steeper, indicating that the ETFs are more sensitive to changes in EQW returns. The confidence intervals in the EQW plots are consistently narrow, showing low uncertainty around the estimated regression line. This implies that the ETFs’ returns are more reliably explained by the EWP.

\newpage

\begin{figure}[H]
    \centering
    \begin{minipage}{0.45\textwidth}
        \centering
        \includegraphics[width=\linewidth]{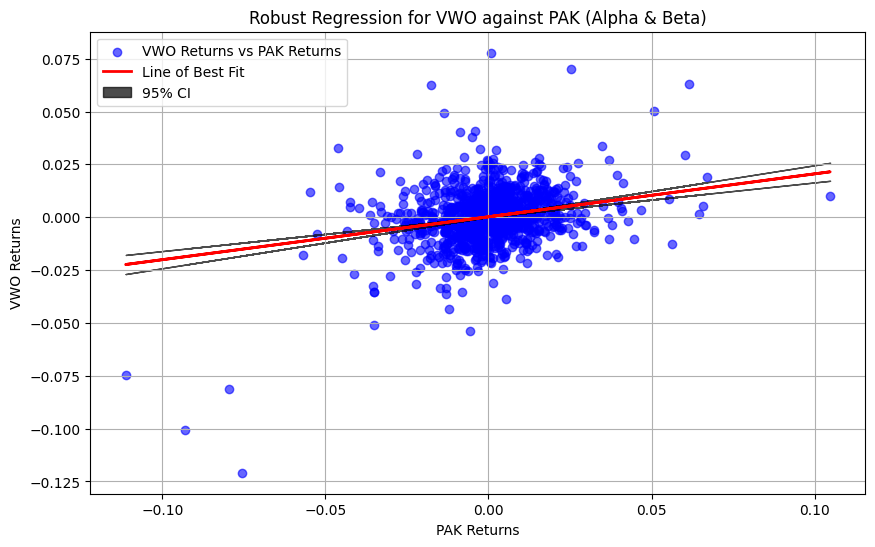}
        \captionsetup{labelformat=empty} 
        \caption{VWO}
    \end{minipage}%
    \hspace{0.05\textwidth}
    \begin{minipage}{0.45\textwidth}
        \centering
        \includegraphics[width=\linewidth]{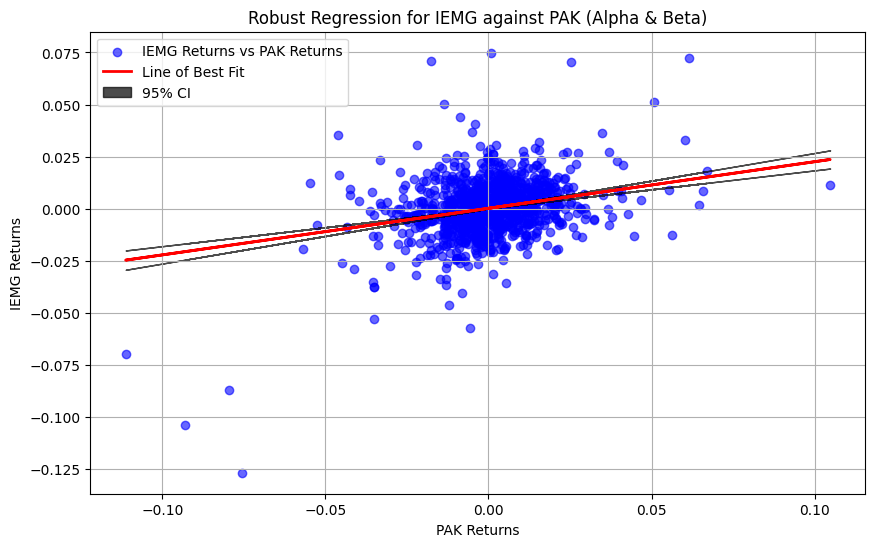}
        \captionsetup{labelformat=empty} 
        \caption{IEMG}
    \end{minipage}
    
    \vspace{0.5cm}
    
    \begin{minipage}{0.45\textwidth}
        \centering
        \includegraphics[width=\linewidth]{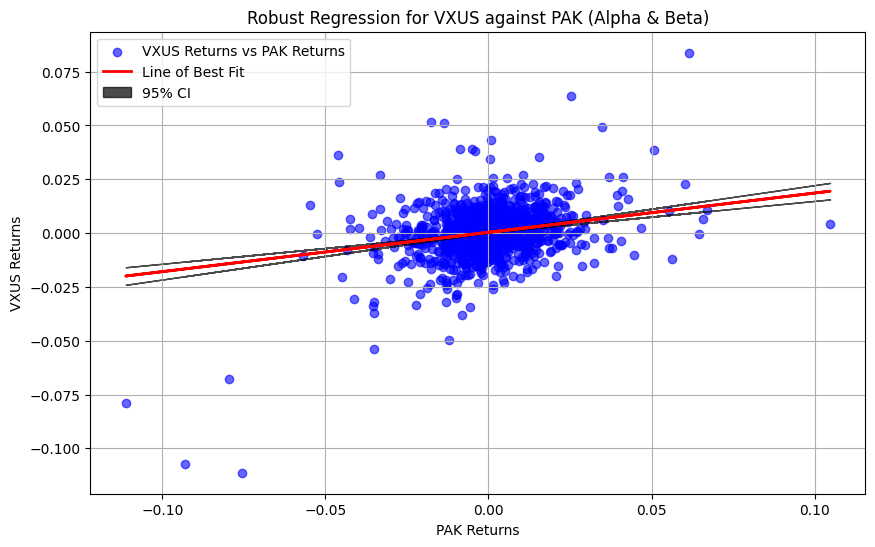}
        \captionsetup{labelformat=empty} 
        \caption{VXUS}
    \end{minipage}%
    \hspace{0.05\textwidth}
    \begin{minipage}{0.45\textwidth}
        \centering
        \includegraphics[width=\linewidth]{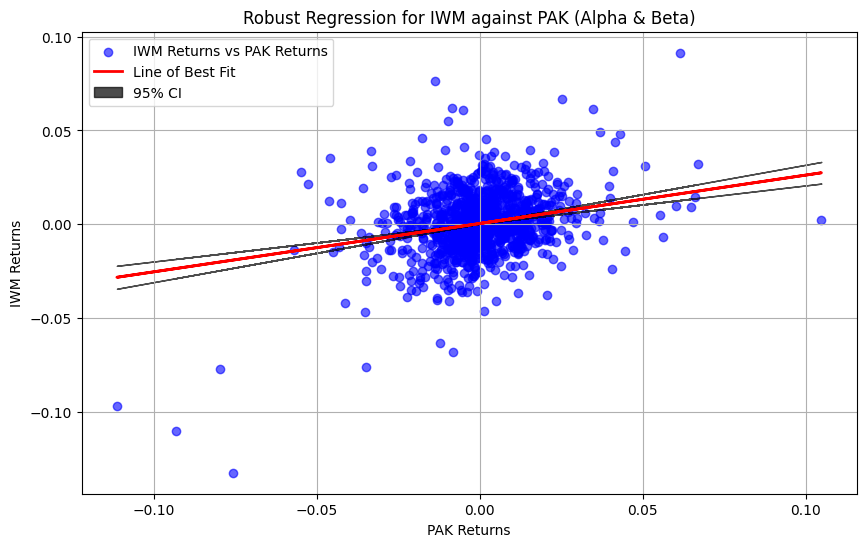}
        \captionsetup{labelformat=empty} 
        \caption{IWM}
    \end{minipage}
    
    \vspace{0.5cm}
    
    \begin{minipage}{0.45\textwidth}
        \centering
        \includegraphics[width=\linewidth]{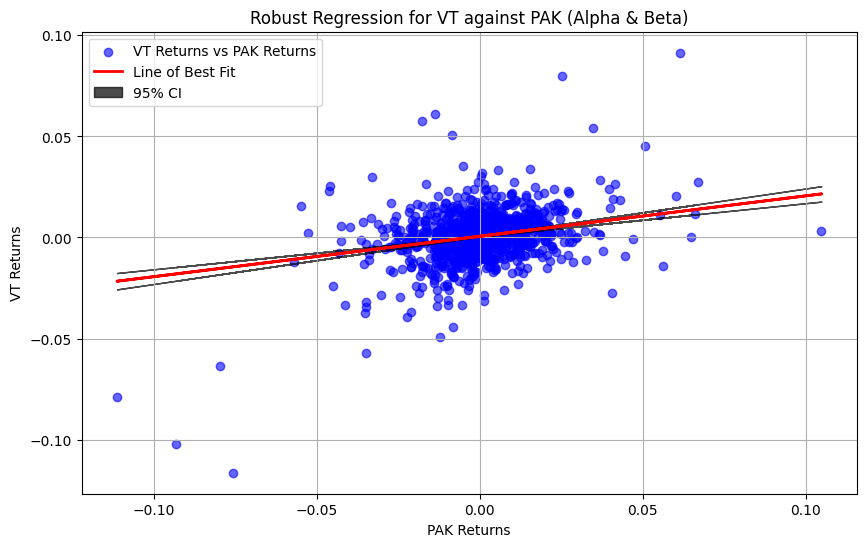}
        \captionsetup{labelformat=empty} 
        \caption{VT}
    \end{minipage}%
    \hspace{0.05\textwidth}
    \begin{minipage}{0.45\textwidth}
        \centering
        \includegraphics[width=\linewidth]{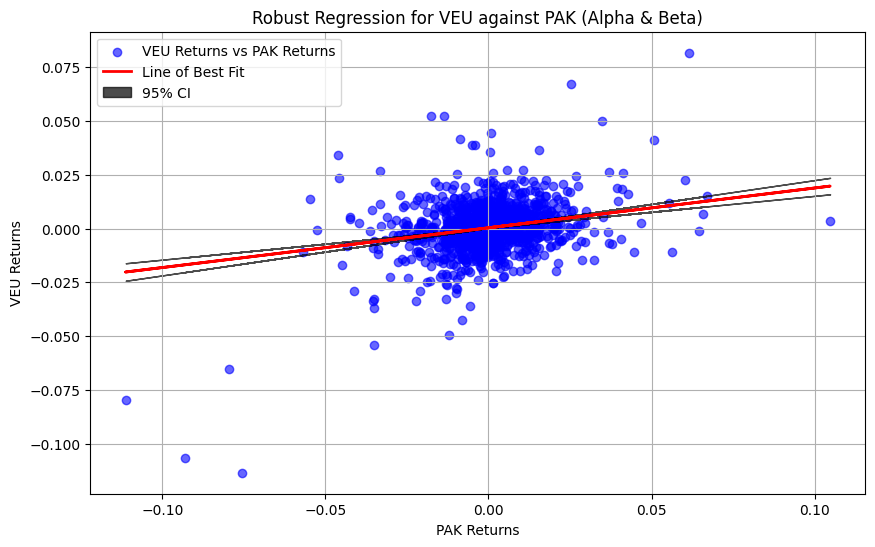}
        \captionsetup{labelformat=empty} 
        \caption{VEU}
    \end{minipage}
    
    \vspace{0.5cm}
    
    \begin{minipage}{0.45\textwidth}
        \centering
        \includegraphics[width=\linewidth]{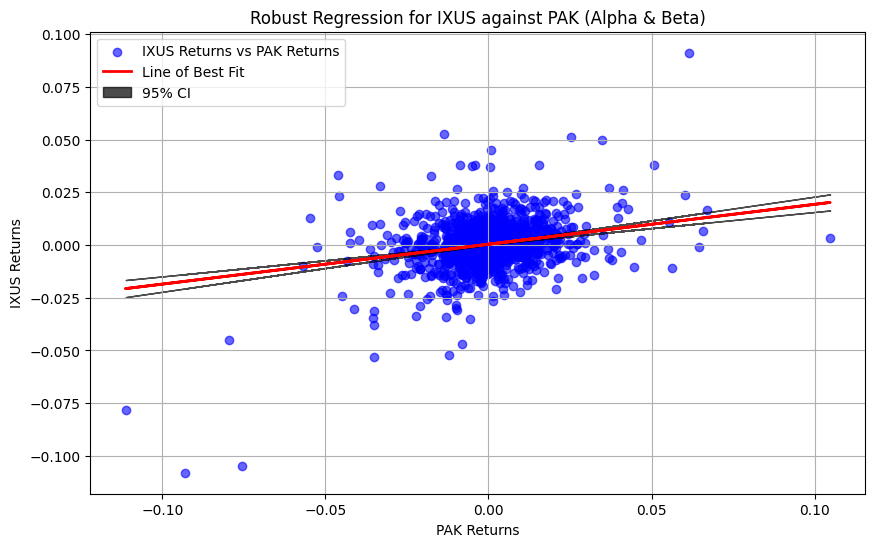}
        \captionsetup{labelformat=empty} 
        \caption{IXUS}
    \end{minipage}%
    \hspace{0.05\textwidth}
    \begin{minipage}{0.45\textwidth}
        \centering
        \includegraphics[width=\linewidth]{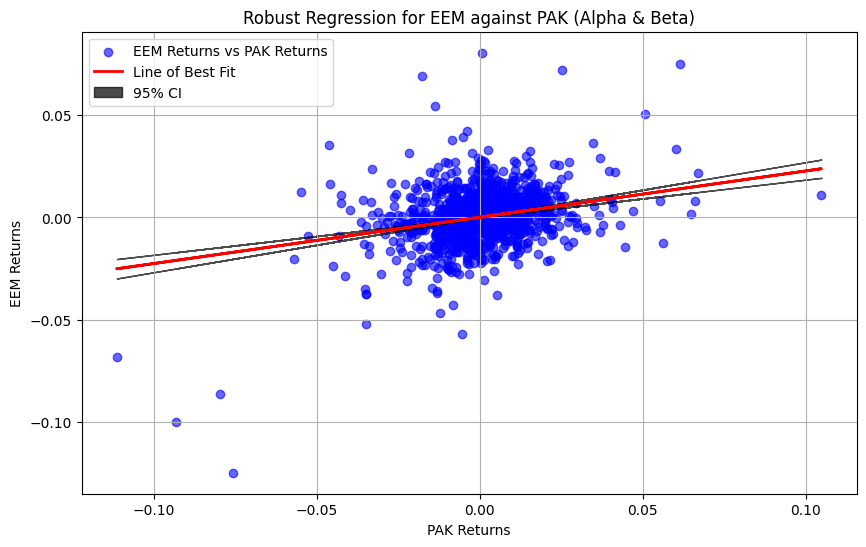}
        \captionsetup{labelformat=empty} 
        \caption{EEM}
    \end{minipage}
\end{figure}

\clearpage

\begin{figure}[H]
    \centering
    \begin{minipage}{0.45\textwidth}
        \centering
        \includegraphics[width=\linewidth]{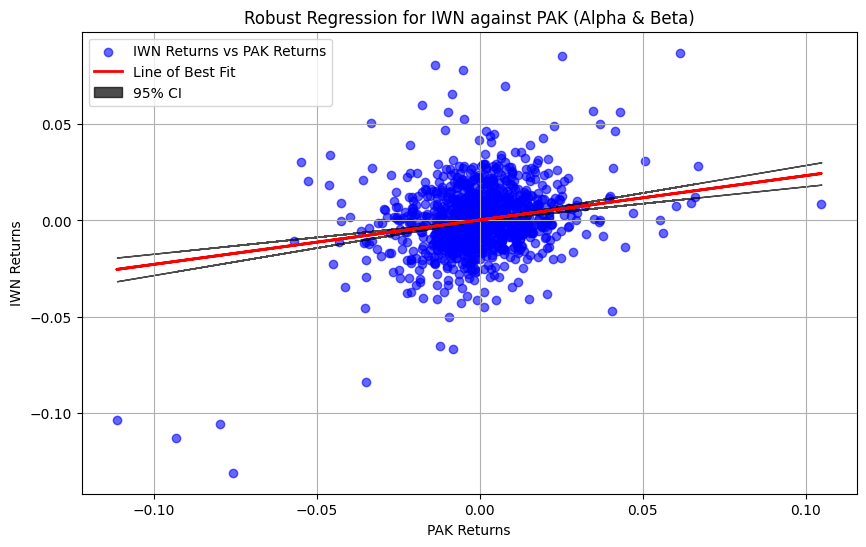}
        \captionsetup{labelformat=empty} 
        \caption{IWN}
    \end{minipage}%
    \hspace{0.05\textwidth}
    \begin{minipage}{0.45\textwidth}
        \centering
        \includegraphics[width=\linewidth]{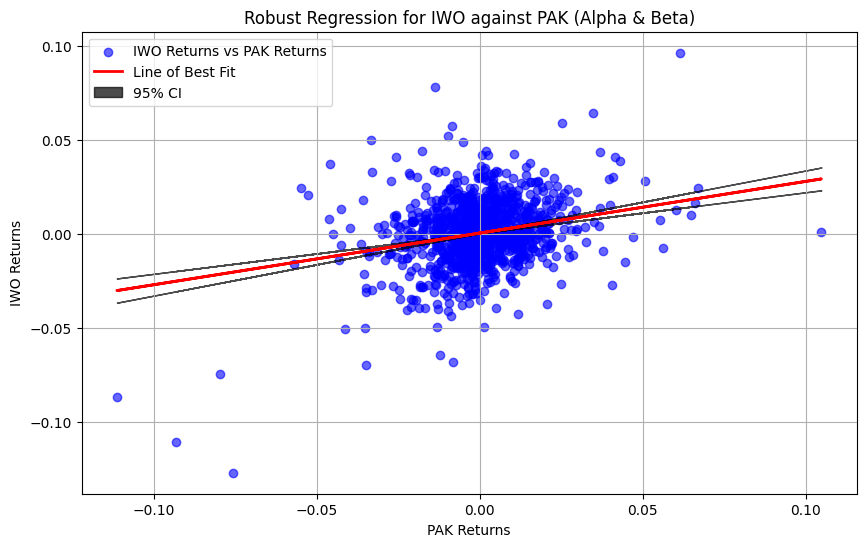}
        \captionsetup{labelformat=empty} 
        \caption{IWO}
    \end{minipage}

\end{figure}

The ETFs generally exhibit a positive association with the PAK benchmark; however, the relationship is weaker than that with the EWP. There is more scatter around the regression line and wider confidence intervals, suggesting greater variability and less predictability in their alignment with PAK. Slopes are generally less steep, indicating a lower sensitivity to PAK returns. This suggests that the ETFs do not respond as strongly or proportionally to changes in PAK returns.

In comparative terms, the EWP benchmark provides a stronger, more predictable fit for the ETFs, with higher sensitivity, tighter clustering and narrower confidence intervals. This implies that the EWP is a more appropriate benchmark for these ETFs, especially those with a large-cap or international focus. The PAK benchmark, while positively correlated, shows a weaker relationship, with more variability and lower sensitivity, suggesting that it may be a less reliable benchmark for these ETFs, particularly for those exposed to broader international markets.

\subsection{Markovitz Efficient Frontier}

In the method introduced by \cite{markowitz}, the objective of portfolio optimization is to determine the set of asset weights that minimize the portfolio return risk, given a desired level of expected portfolio return $\Tilde{r}_{p}$. The target value of $\Tilde{r}_{p}$ reflects the investor's risk tolerance; a higher $\Tilde{r}_{p}$ indicates a greater willingness to accept risk. Using the portfolio variance $\sigma_{p}^2$ as a measure of risk, Markowitz’s mean-variance optimization framework seeks to minimize this variance, subject to constraints on the expected return and ensuring full investment across all assets. Formally, this can be expressed as the minimization of the portfolio variance under the constraints of the desired expected return and total allocation of investment capital.

We follow the Markowitz mean-variance portfolio optimization problem, as outlined in \citet{reit-portfolio}. The optimization problem is solved using standard methods of employing Lagrange multipliers:

\begin{equation}
    \min_{w, q, \theta_{0}}\;L(w, q, \theta_{0}) =  \min_{w, q, \theta_{0}}(\frac{w^{T}\Sigma{w}}{2} + q(\Bar{r}_{p} - \Bar{r}^{T}w) + \theta_{0}(1 - e_{n}^{T}w)).
\end{equation}

\noindent Taking the first-order conditions yields the following optimality conditions for $w$:

\begin{equation}
    w^* = \Bar{r}_{p}w_{1} + w_{2}, 
\end{equation}

\noindent and the variance of the portfolio is given by

\begin{equation*}
   \sigma_{p} = \sqrt{w^{*T} \Sigma w^*} = \sqrt{\frac{B\Bar{r}_{p}^2 - 2C\Bar{r}_{p} + A}{\Delta}}
\end{equation*}

\noindent where $A = \Bar{r}^{T}\Sigma^{-1}\Bar{r}$, $B = e_{n}^{T}\Sigma^{-1}e_{n}$ and $C = \Bar{r}^{T}\Sigma^{-1}e_{n}$. Therefore, $\Delta = AB - C^2$. Given these relationships, $(\sigma_{p}(w^*), \Bar{r}_{p})$ are the portfolio frontier points that trace out a hyperbola in the risk (standard deviation) and return (mean) space.  

The following plot shows the Markowitz efficient frontier for a set of portfolios composed of various assets, including the EWP and PAK ETFs, along with other ETFs. The efficient frontier represents the set of portfolios that offer the highest expected return for a given level of risk or the lowest risk for a given level of return. The capital market line (CML) represents the risk-return trade-off for portfolios that combine a risk-free asset with the market portfolio. The point where the CML intersects the efficient frontier represents the tangency portfolio, or the optimal market portfolio, which maximizes the Sharpe ratio. Each dot represents the risk and return profile of a single ETF. The scattered positions indicate a variety of risk-return trade-offs across the ETF options. The EWP is closer to the efficient frontier, making it a relatively efficient choice among the ETFs, providing a good risk-return balance. PAK lies below the efficient frontier, indicating that it is an inefficient choice in this risk-return space; investors are exposed to more risk than necessary for the return PAK offers.

\begin{figure}[H]
\captionsetup{justification=centering}
\centering
\includegraphics[width=1.0\textwidth]{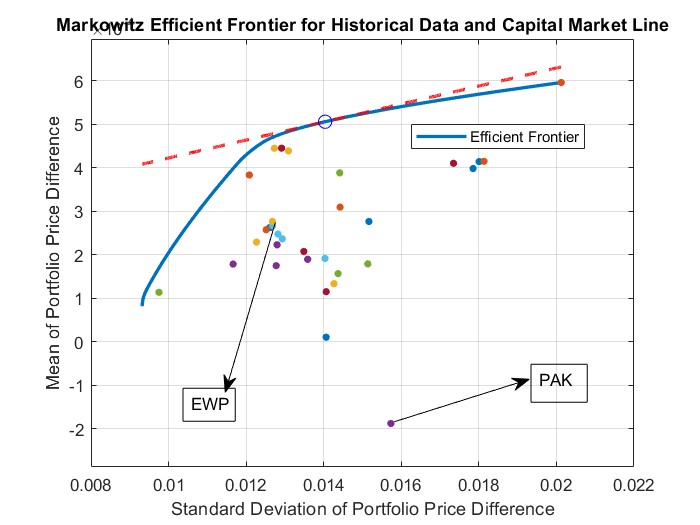}
\caption[The system.]{Markowitz efficient frontier (historical optimization).}
\label{theSystemFig}
\end{figure}

\subsection{Key Performance Ratios}

Table 4 illustrates the Sharpe ratios for various portfolios; they measure risk-adjusted returns. Among these, long-short TC95 and TC99 exhibit the highest Sharpe ratio of 0.0595, indicating that they are the most efficient portfolios in terms of returns per unit of risk. Long-short C95 follows with a Sharpe ratio of 0.0289, showing a moderate performance. Long-only TVP, TC95 and TC99 have identical Sharpe ratios of 0.0226, suggesting a similar performance but with a lower efficiency compared to the top-performing portfolios. Notably, long-only (LO) C95 has a negative Sharpe ratio of -0.0010, implying that it incurs losses relative to its risk. Portfolios like the LO MVP and EWP have minimal positive Sharpe ratios, suggesting that they offer low risk-adjusted returns, making them less attractive for risk-averse investors. Overall, the LS portfolios, particularly the top three, outperform the LO portfolios in terms of the risk-adjusted performance.

\begin{table}[H]
\begin{center}
    \begin{tabular}{|l|c|}
        \hline
        \textbf{Portfolio} & \textbf{Sharpe Ratio} \\ \hline
        LS\_TVP & 0.059474467 \\ \hline
        LS\_TC95 & 0.059474467 \\ \hline
        LS\_TC99 & 0.059474467 \\ \hline
        LS\_C95 & 0.028857923 \\ \hline
        LO\_TVP & 0.022564225 \\ \hline
        LO\_TC95 & 0.022564225 \\ \hline
        LO\_TC99 & 0.022564225 \\ \hline
        LS\_MVP & 0.01856749 \\ \hline
        LS\_C99 & 0.012710158 \\ \hline
        EWP & 0.007972721 \\ \hline
        LO\_C99 & 0.004834784 \\ \hline
        LO\_MVP & 0.000401662 \\ \hline
        LO\_C95 & -0.001006328 \\ \hline
    \end{tabular}
    \caption{Sharpe Ratios of Portfolios}
    \label{tab:section3_5_sharpe_ratio}
\end{center}
\end{table}

\begin{table}[H]
\begin{center}
    \begin{tabular}{|l|c|}
        \hline
        \textbf{Portfolio} & \textbf{Max. Drawdown} \\ \hline
        LS\_TC95 & 0.591018918 \\ \hline
        LS\_TC99 & 0.591018918 \\ \hline
        LS\_TVP & 0.591018918 \\ \hline
        EWP & 0.365943773 \\ \hline
        LO\_TVP & 0.317027745 \\ \hline
        LO\_TC95 & 0.317027745 \\ \hline
        LO\_TC99 & 0.317027745 \\ \hline
        LO\_MVP & 0.313304961 \\ \hline
        LO\_C99 & 0.308339959 \\ \hline
        LO\_C95 & 0.303296926 \\ \hline
        LS\_MVP & 0.239419095 \\ \hline
        LS\_C95 & 0.191313723 \\ \hline
        LS\_C99 & 0.158217606 \\ \hline
    \end{tabular}
    \caption{Maximum Drawdowns of Portfolios}
    \label{tab:section3_5_max_drawdown}
\end{center}
\end{table}

\newpage
Table 5 shows the maximum drawdowns (MDDs) of various portfolios, with the highest MDD of 0.591 observed in long-short TC95, TC99 and TVP, indicating that these portfolios are the most vulnerable to peak-to-trough declines, likely due to aggressive strategies or higher risk exposure. In contrast, long-short C99 exhibits the lowest MDD at 0.158, showcasing superior resilience and effective risk control. Portfolios like the EWP  (MDD 0.366) and the long-only portfolios (MDD ranging from 0.317 to 0.303) offer moderate risk profiles, balancing returns and vulnerability.


\begin{table}[H]
\begin{center}
    \begin{tabular}{|l|c|}
        \hline
        \textbf{Portfolio} & \textbf{Calmar Ratio} \\ \hline
        LS TVP & 1.486 \\ \hline
        LS TC95 & 1.486 \\ \hline
        LS TC99 & 1.486 \\ \hline
        LS C95 & 0.406 \\ \hline
        LO TVP & 0.365 \\ \hline
        LO TC95 & 0.365 \\ \hline
        LO TC99 & 0.365 \\ \hline
        LS C99 & 0.326 \\ \hline
        LS MVP & 0.237 \\ \hline
        EWP & 0.138 \\ \hline
        LO C99 & 0.118 \\ \hline
        LO MVP & 0.079 \\ \hline
        LO C95 & 0.070 \\ \hline
    \end{tabular}
    \caption{Calmar Ratios of LS and LO Portfolios}
    \label{tab:calmar_ratios}
\end{center}
\end{table}

The long-short (LS) portfolios dominate in terms of performance, with the TVP, TC95 and TC99 achieving the highest Calmar ratio of 1.486, indicating that these portfolios generate the highest return relative to their drawdown risk. This suggests that they are exceptionally efficient in balancing returns and risks. Among the remaining LS portfolios, C95 has a Calmar ratio of 0.406, followed by C99 at 0.326 and the MVP at 0.237, showing a gradual decline in efficiency.

The long-only (LO) portfolios exhibit significantly lower Calmar ratios, with the TVP, TC95 and TC99 clustered at 0.365, suggesting moderate performance. Other LO portfolios, such as C99 (0.118) and C95 (0.070), have the lowest Calmar ratios, indicating less efficient risk-adjusted returns. The EWP, with a Calmar ratio of 0.138, performs slightly better than the least efficient LO portfolios but remains significantly below the top-performing LS portfolios. This analysis highlights that LS strategies, particularly the TVP, TC95 and TC99, provide superior risk-adjusted returns compared to LO portfolios.

\begin{table}[H]
\begin{center}
    \begin{tabular}{|l|c|}
        \hline
        \textbf{Portfolio} & \textbf{STARR} \\ \hline
        LS TVP & 1.059 \\ \hline
        LS TC95 & 1.059 \\ \hline
        LS TC99 & 1.059 \\ \hline
        LS C95 & 0.622 \\ \hline
        LS MVP & 0.440 \\ \hline
        LO TVP & 0.361 \\ \hline
        LO TC95 & 0.361 \\ \hline
        LO TC99 & 0.361 \\ \hline
        LS C99 & 0.257 \\ \hline
        EWP & 0.165 \\ \hline
        LO C99 & 0.141 \\ \hline
        LO MVP & 0.097 \\ \hline
        LO C95 & 0.083 \\ \hline
    \end{tabular}
    \caption{STARR Values of LS and LO Portfolios}
    \label{tab:starr_ratios}
\end{center}
\end{table}

Table 7 presents the stable tail adjusted return ratio (STARR) values for various portfolios; they measure the risk-adjusted performance considering the tail risk. The LS portfolios consistently outperform the LO portfolios. LS TVP, LS TC95 and LS TC99 achieve the highest STARR value of 1.059, showcasing a superior tail-risk-adjusted performance. LS C95 follows with a STARR of 0.622, and LS MVP has a STARR of 0.440, with both maintaining a clear advantage over LO portfolios. Among the LO portfolios, the TVP, TC95 and TC99 exhibit identical STARR values of 0.361, representing moderate performance. The lowest-performing portfolios include LO C99 (0.141), LO MVP (0.097) and LO C95 (0.083), indicating weaker returns relative to their tail risks. The EWP sits between the LS and LO portfolios with a STARR of 0.165. This analysis highlights that LS strategies, particularly the TVP, TC95 and TC99, are optimal for tail-risk-conscious investors, while LO portfolios show a lower risk-adjusted efficiency.

\section{Dynamic Portfolio Optimization}

The historical optimization approach, as described  and further demonstrated in the previous section, involves sequentially sampling return data using a rolling window technique over a fixed historical period that captures a finite range of market conditions. However, as is often highlighted in fund prospectuses, historical performance does not necessarily predict future outcomes. Instead of relying solely on historical asset return samples, dynamic optimization aims to enhance the insights that can be extracted from historical data. This approach assumes that historical returns originate from a dynamic multivariate distribution—dynamic in the sense that its statistical properties, such as covariance, may evolve over time. Dynamic optimization focuses on characterizing this distribution and generating extensive predictive samples of correlated asset returns, specifically aiming to capture more of the distribution's tail behavior, including extreme events. The outcome is a portfolio optimization process that is better calibrated to anticipate significant shifts in market performance \citep{reit-portfolio}. 

In this section, we implement dynamic optimization. There are some key parts of this optimization. In each rolling window, we will fit a common time series model, the ARMA(1,1)-GARCH(1,1) model, which will be referred to as the AG model,  to the return data of the ETFs. In addition to that, a Student's t distribution that takes into account extreme scenarios will be employed as an empirical model for the innovations in the AG fit. The process begins by transforming the data into a format where all parts of the distribution, including extreme values (tails), are treated equally. This is done using a ``copula transformation." After the transformation, the data are modeled using a multivariate copula, which captures how the assets move together (their correlations). Using this model, a large set of possible values for the asset movements is simulated. These simulated values are then converted back into their original format using an inverse transformation. Finally, these values are used to estimate a wide range of potential portfolio returns, which are then fed into the optimization process to determine the best asset weights for the next day. The main purpose of this optimization  is to come up with a statistically correct large-sample process. We will achieve this in a way that uses the initial historical window of return data with a dynamic forecast of returns for the next day that will be utilized in the optimization for determining weights on day t+1.

\newpage
\subsection{AG Model with Student's t Distribution}
We will use the empirical specification for the ARMA(p,q) model developed by \cite{tsay2005analysis}, and the ARMA(1,1)-GARCH(1,1) model will be defined as follows:

\begin{equation*}
    r_{t} = \delta_{0} + \sum_{i=1}^{p} \delta_{i} r_{t-i} + \alpha_{t} + \sum_{j=1}^{q} \delta_{j} \alpha_{j-1},
\end{equation*}
\begin{center}
    $\alpha_{t} = \sigma_{t} \epsilon_{t}$,
\end{center}
\begin{equation*}
    \sigma_t^2 = \alpha_0 + \alpha_1 a_{t-1}^2 + \beta_1 \sigma_{t-1}^2,
\end{equation*}

where $\alpha_{t}$ is a shock. In this specification, our assumption is that the residuals $\epsilon_{t}$ follow a Student's t distribution:

\begin{equation}
    t_{\nu} (x) = \frac{\Gamma\left(\frac{1+\nu}{2}\right)}{\sqrt{\nu \pi}  \Gamma \left(\frac{\nu}{2}\right)} 
    \left(1 + \frac{x^2}{\nu}\right)^{-\frac{1+\nu}{2}}.
\end{equation}

where $\Gamma$() denotes the gamma function. The subject distribution is symmetric but relatively  leptokurtic in comparison to the normal distribution. As shown in Figure 4.1, a window of historical returns is transformed into a dynamic set of returns, which are then passed to the portfolio-optimizing routine.

\begin{figure}[H]
\captionsetup{justification=centering}
\centering
\includegraphics[width=1.0\textwidth]{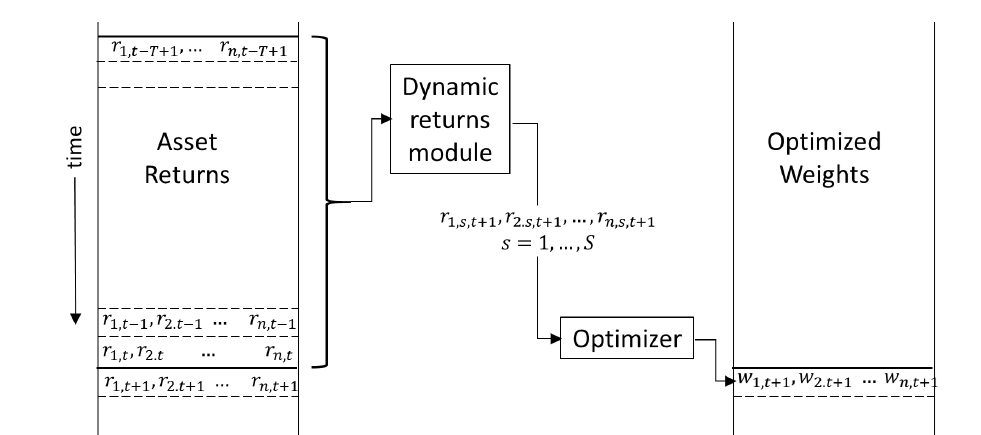}
\caption[The system.]{Schematic of dynamic portfolio optimization \citep{reit-portfolio}.}
\label{theSystemFig}
\end{figure}


\subsection{Basic Strategies, Price and Return Performance}
\subsubsection{Long Only }

Figure 4.2 illustrates the performance of various portfolio strategies, starting with a \$100 investment, from early 2020 to 2024. The EWP outperforms all other strategies, but it exhibits higher volatility, making it more suitable for risk-tolerant investors. Portfolios such as the MVP and TVP  show moderate returns with lower volatility, offering a balanced approach for investors seeking steady growth. Overall, the EWP provides the best returns for aggressive investors, while the MVP and TVP are better suited for those with moderate risk tolerance. Conservative investors may need to reconsider  TC99, as it fails to deliver adequate returns.

\begin{figure}[H]
\captionsetup{justification=centering}
\centering
\includegraphics[width=1.0\textwidth]{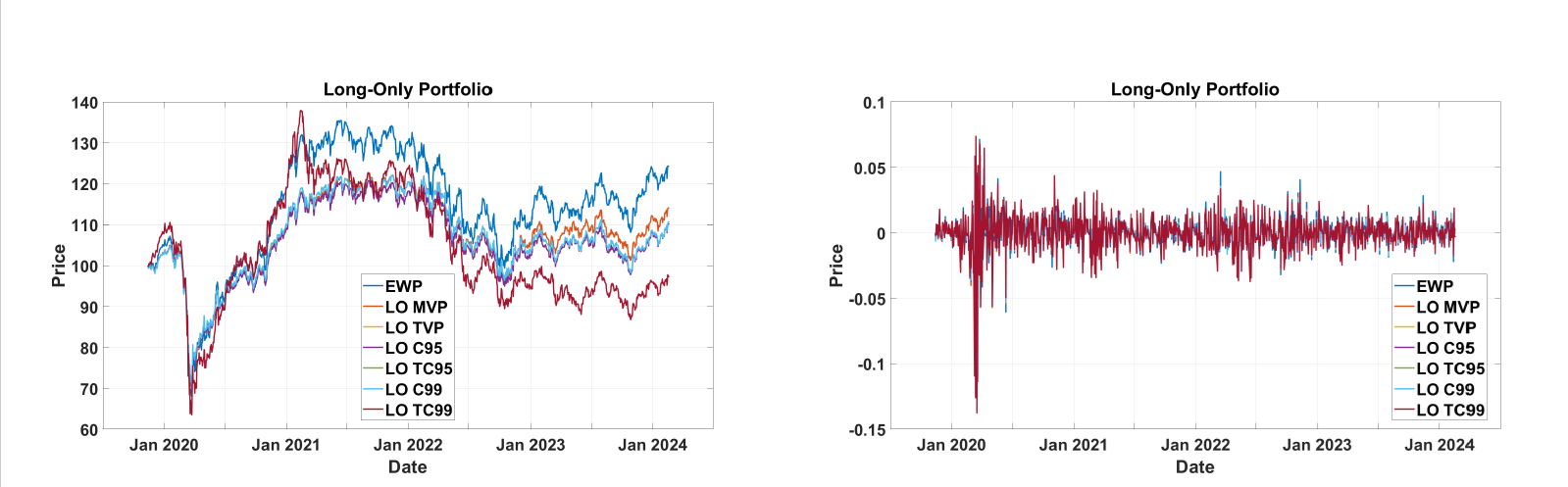}
\caption[The system.]{Comparison of the cumulative price (left) and log-return (right) of the long-only portfolios to those of the benchmark.}
\label{theSystemFig}
\end{figure}

\subsubsection{Long Short}

The performance comparison of long-short portfolios reveals significant differences in risk and return dynamics. The TC99 strategy emerges as the top performer, with aggressive growth, but it also exhibits high volatility, making it suitable for risk-tolerant investors. In contrast, conservative strategies like LS MVP and LS TVP demonstrate steady performance with low volatility, appealing to risk-averse investors. Portfolios such as LS C95 and LS TC95 offer a balance between risk and return, outperforming LS MVP and LS TVP while maintaining moderate volatility. Overall, LS TC99 is ideal for aggressive growth, while LS MVP and LS TVP cater to investors prioritizing stability.

\begin{figure}[H]
\captionsetup{justification=centering}
\centering
\includegraphics[width=1.0\textwidth]{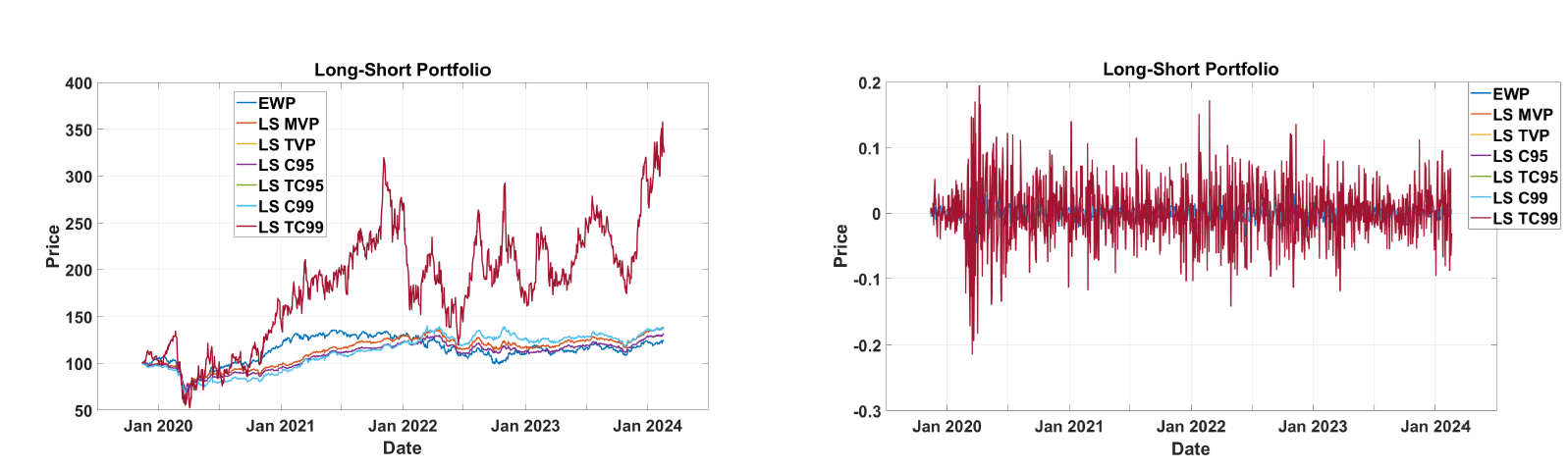}
\caption[The system.]{Comparison of the cumulative price (left) and log-return (right) of the long-short portfolios to those of the benchmark.}
\label{theSystemFig}
\end{figure}

\subsection{Efficient Frontier}
The Markowitz efficient frontier based on the dynamic optimization of 30 ETFs in Figure 4.4 shows the trade-off between risk (standard deviation of portfolio price differences) and return (mean of portfolio price differences). The efficient frontier (blue curve) highlights the portfolios that maximize returns for a given risk, while the CML indicates the optimal risk-return combinations when incorporating a risk-free asset. The EWP lies slightly below the efficient frontier, offering a balanced but suboptimal risk-return profile. In contrast, the Pakistan ETF (PAK) is inefficient, with higher risk and lower returns, demonstrating a poor risk-adjusted performance. The tangency point on the CML represents the market portfolio, offering the best possible risk-return trade-off. This analysis underscores the importance of diversification and optimization in portfolio construction to achieve superior risk-adjusted returns.
 
\begin{figure}[H]
\captionsetup{justification=centering}
\centering
\includegraphics[width=1.0\textwidth]{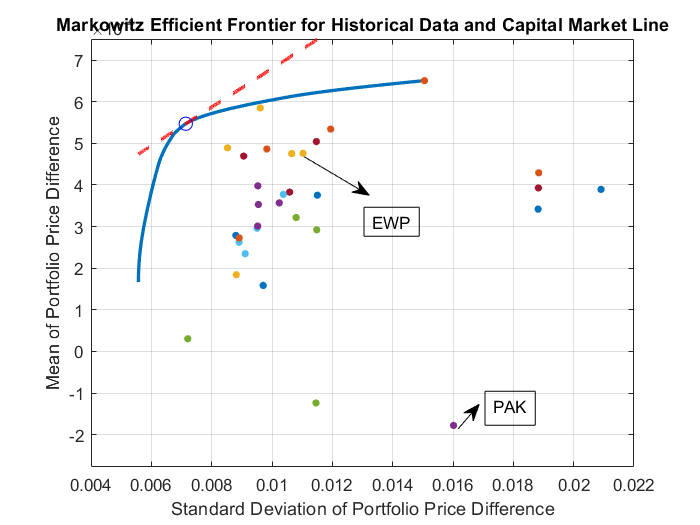}
\caption[The system.]{Markowitz efficient frontier (dynamic optimization).}
\label{theSystemFig}
\end{figure}

\section{Dow Jones Industrial Average}

\subsection{Basic Strategies, Price and Return Performance}
\subsubsection{Long Only }
\justifying  The performance of the cumulative price of each portfolio with respect to the Dow Jones from 11/13/2019 through 2/19/2024, assuming a \$100 investment in the portfolio on 11/12/2019, is shown in Figure 5.1. The portfolio TC99 exhibits exponential growth, far outperforming all other portfolios. This strategy, however, also displays substantial volatility. On the other hand, conservative portfolios such as LO MVP and LO TVP  demonstrate a stable, low-risk performance, with minimal price growth and less variability, making them suitable for risk-averse investors. The EWP, which is the DJIA index, provides modest returns, performing better than LO MVP and LO TVP but significantly underperforming compared to LO TC99. Portfolios like C95 and TC95 strike a middle ground, achieving moderate growth with balanced risk. Overall, TC99 provides exceptional returns but at the cost of high risk, whereas the MVP and  TVP cater to investors prioritizing stability over growth.

\begin{figure}[H]
\captionsetup{justification=centering}
\centering
\includegraphics[width=1.0\textwidth]{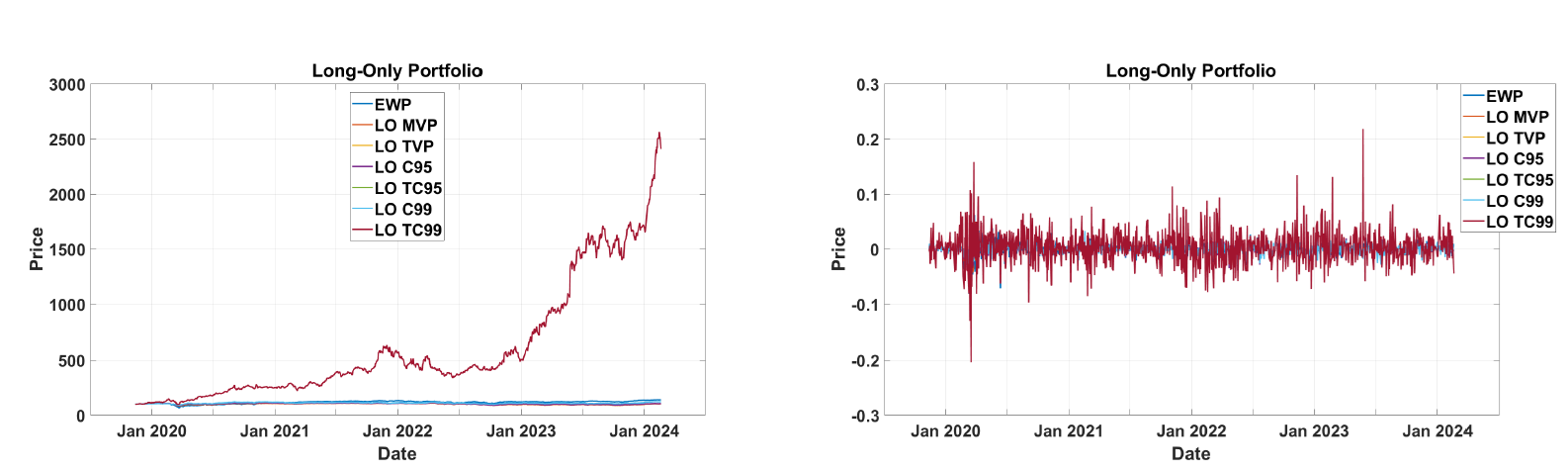}
\caption[The system.]{Comparison of the cumulative price (left) and log-return (right) of the long-only portfolios to those of the benchmark.}
\label{theSystemFig}
\end{figure}

\subsubsection{Long Short}

The analysis of long-short portfolios for the DJIA with an initial investment of \$100 highlights notable differences in performance and risk profiles. The LS TC99 strategy demonstrates exceptional growth. This exponential growth, however, comes with significant risk, as seen in the high volatility reflected in the right panel. Other portfolios, such as LS C95 and LS TC95, also deliver impressive returns but with slightly lower growth trajectories compared to LS TC99. The EWP, LS MVP and LS TVP  exhibit comparatively low returns and minimal volatility, making them better suited for risk-averse investors. The right panel further emphasizes that strategies like TC99 and  C95 experience the highest levels of volatility, while the EWP and LS MVP remain stable. Overall, TC99 stands out for its exceptional return potential but involves significant risk, while the EWP, MVP and TVP provide stable but subdued growth for conservative investors.

\begin{figure}[H]
\captionsetup{justification=centering}
\centering
\includegraphics[width=1.0\textwidth]{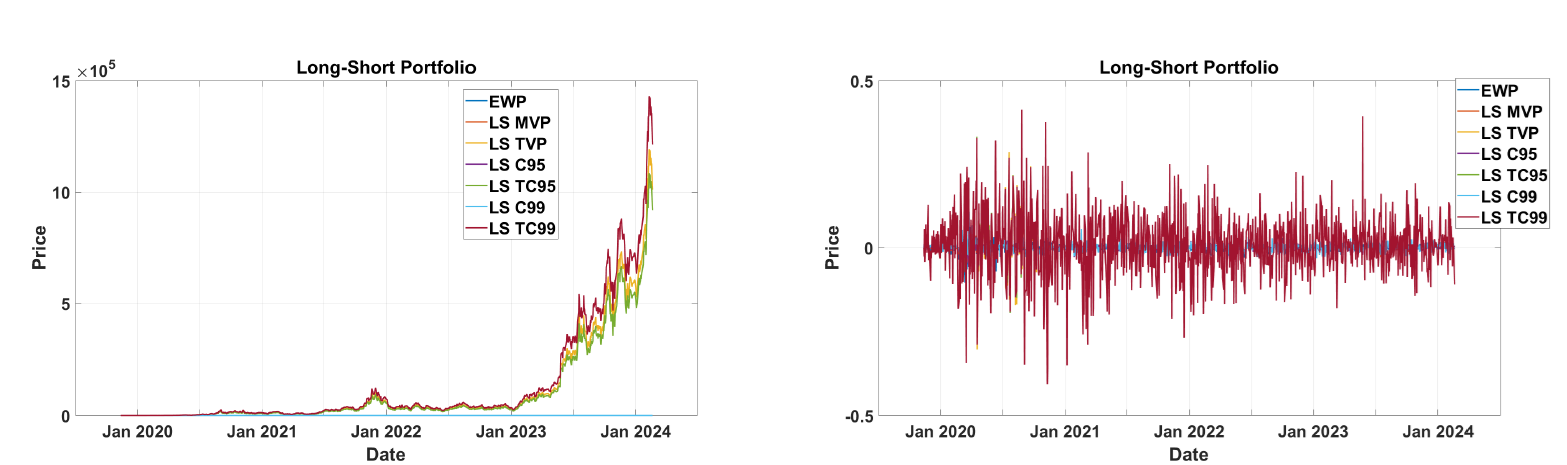}
\caption[The system.]{Comparison of the cumulative price (left) and log-return (right) of the long-short portfolios to those of the benchmark.}
\label{theSystemFig}
\end{figure}

\subsection{Efficient Frontier}

\begin{figure}[H]
\captionsetup{justification=centering}
\centering
\includegraphics[width=1.0\textwidth]{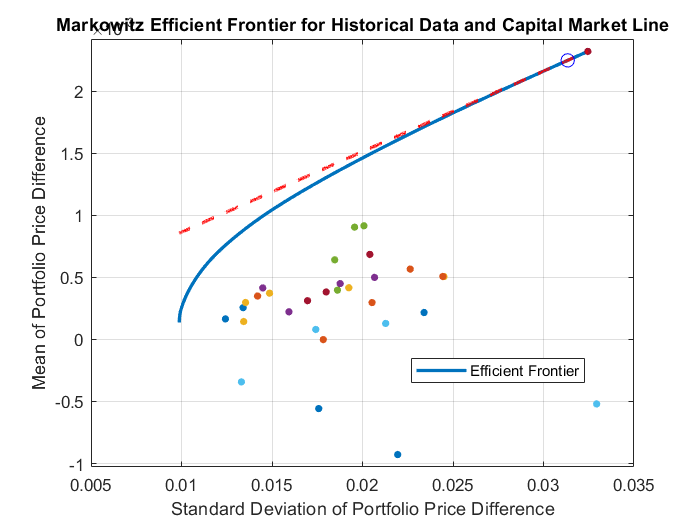}
\caption[The system.]{Markowitz efficient frontier for DJIA (historical optimization).}
\label{theSystemFig}
\end{figure}

\newpage

\section{Conclusions}

The analysis of Pakistan-exposed ETFs highlights both the opportunities and challenges of investing in frontier and emerging markets. While Pakistan’s market exhibits unique characteristics, such as a low correlation with global markets and potential for diversification, the performance of ETFs like PAK reveals inefficiencies. Positioned below the efficient frontier, PAK offers lower returns for higher risk, making it suboptimal compared to diversified benchmarks like the EWP. However, the integration of Pakistan-focused ETFs into broader portfolios underscores their potential to mitigate risk through diversification, particularly for US investors seeking exposure to non-traditional markets. The findings demonstrate that while Pakistan's structural challenges—such as political instability and market illiquidity—limit performance, the dynamic optimization framework offers pathways to enhance portfolio efficiency by capturing tail risks and adapting to evolving market conditions.

Dynamic optimization strategies, particularly those incorporating ARMA-GARCH models and Student's t distributions, outperform historical approaches by effectively modeling risk and return trade-offs. Long-short strategies such as LS TC99 deliver superior returns, albeit with significant volatility, making them ideal for risk-tolerant investors, while conservative strategies like LS MVP and LS TVP cater to risk-averse preferences. The study underscores the critical role of advanced optimization techniques in managing the complexities of frontier markets, where traditional methodologies often fail to capture extreme events or evolving correlations. Overall, this research not only provides actionable insights for investors considering Pakistan-exposed ETFs but also contributes to the broader discourse on portfolio optimization in high-risk, high-reward markets.

\newpage
\section{Appendix}
\subsection{Historical Performance Ratios of ETFs}

\begin{table}[H]
\begin{center}
    \begin{tabular}{|l|c|c|c|c|}
        \hline
        \textbf{Portfolio} & \textbf{VaR\_95} & \textbf{CVaR\_95} & \textbf{VaR\_99} & \textbf{CVaR\_99} \\ \hline
        LO MVP & -0.0128 & -0.0230 & -0.0273 & -0.0487 \\ \hline
        LO TVP & -0.0238 & -0.0371 & -0.0432 & -0.0624 \\ \hline
        LO C95 & -0.0125 & -0.0227 & -0.0247 & -0.0493 \\ \hline
        LO TC95 & -0.0238 & -0.0371 & -0.0432 & -0.0624 \\ \hline
        LO C99 & -0.0137 & -0.0234 & -0.0272 & -0.0481 \\ \hline
        LO TC99 & -0.0238 & -0.0371 & -0.0432 & -0.0624 \\ \hline
        LS MVP & -0.0091 & -0.0157 & -0.0175 & -0.0325 \\ \hline
        LS TVP & -0.0647 & -0.0923 & -0.1114 & -0.1259 \\ \hline
        LS C95 & -0.0091 & -0.0146 & -0.0140 & -0.0304 \\ \hline
        LS TC95 & -0.0647 & -0.0923 & -0.1114 & -0.1259 \\ \hline
        LS C99 & -0.0124 & -0.0176 & -0.0189 & -0.0265 \\ \hline
        LS TC99 & -0.0647 & -0.0923 & -0.1114 & -0.1259 \\ \hline
        EWP & -0.0172 & -0.0304 & -0.0334 & -0.0633 \\ \hline
    \end{tabular}
    \caption{VaR and CVaR at 95\% and 99\% Confidence Levels for LS and LO Portfolios}
    \label{tab:var_cvar}
\end{center}
\end{table}

\begin{table}[H]
\begin{center}
    \begin{tabular}{|l|c|}
        \hline
        \textbf{Portfolio} & \textbf{Rachev Ratio} \\ \hline
        LS TC95 & -0.8499 \\ \hline
        LS TC99 & -0.8499 \\ \hline
        LS TVP & -0.8499 \\ \hline
        LS C95 & -0.9166 \\ \hline
        LO TVP & -0.9386 \\ \hline
        LO TC95 & -0.9386 \\ \hline
        LO TC99 & -0.9386 \\ \hline
        LS MVP & -0.9450 \\ \hline
        LS C99 & -0.9659 \\ \hline
        EWP & -0.9763 \\ \hline
        LO C99 & -0.9855 \\ \hline
        LO MVP & -0.9988 \\ \hline
        LO C95 & -1.0031 \\ \hline
    \end{tabular}
    \caption{Rachev Ratios of LS and LO Portfolios}
    \label{tab:rachev_ratios}
\end{center}
\end{table}

\begin{table}[H]
\begin{center}
    \begin{tabular}{|l|c|}
        \hline
        \textbf{Portfolio} & \textbf{Sortino Ratio} \\ \hline
        LS TVP & 0.0599 \\ \hline
        LS TC95 & 0.0599 \\ \hline
        LS TC99 & 0.0599 \\ \hline
        LS C95 & 0.0265 \\ \hline
        LO TVP & 0.0208 \\ \hline
        LO TC95 & 0.0208 \\ \hline
        LO TC99 & 0.0208 \\ \hline
        LS MVP & 0.0165 \\ \hline
        LS C99 & 0.0128 \\ \hline
        EWP & 0.0073 \\ \hline
        LO C99 & 0.0045 \\ \hline
        LO MVP & 0.0004 \\ \hline
        LO C95 & -0.0009 \\ \hline
    \end{tabular}
    \caption{Sortino Ratios of LS and LO Portfolios}
    \label{tab:sortino_ratios}
\end{center}
\end{table}

\begin{table}[H]
\begin{center}
    \begin{tabular}{|l|c|}
        \hline
        \textbf{Portfolio} & \textbf{Jensen's Alpha} \\ \hline
        LS TC95 & 0.0022 \\ \hline
        LS TC99 & 0.0022 \\ \hline
        LS TVP & 0.0022 \\ \hline
        LO TVP & 0.0001 \\ \hline
        LO TC95 & 0.0001 \\ \hline
        LO TC99 & 0.0001 \\ \hline
        LS C95 & 0.0001 \\ \hline
        LS MVP & 0.00006 \\ \hline
        LS C99 & 0.00005 \\ \hline
        EWP & -0.0001 \\ \hline
        LO C99 & -0.0001 \\ \hline
        LO MVP & -0.0001 \\ \hline
        LO C95 & -0.0002 \\ \hline
    \end{tabular}
    \caption{Jensen's Alpha of LS and LO Portfolios}
    \label{tab:jensens_alpha}
\end{center}
\end{table}

\newpage
\subsection{Historical vs. Dynamic ETF Optimization Performance Ratio Comparison}

\begin{figure}[H]
\captionsetup{justification=centering}
\centering
\includegraphics[width=1.0\textwidth]{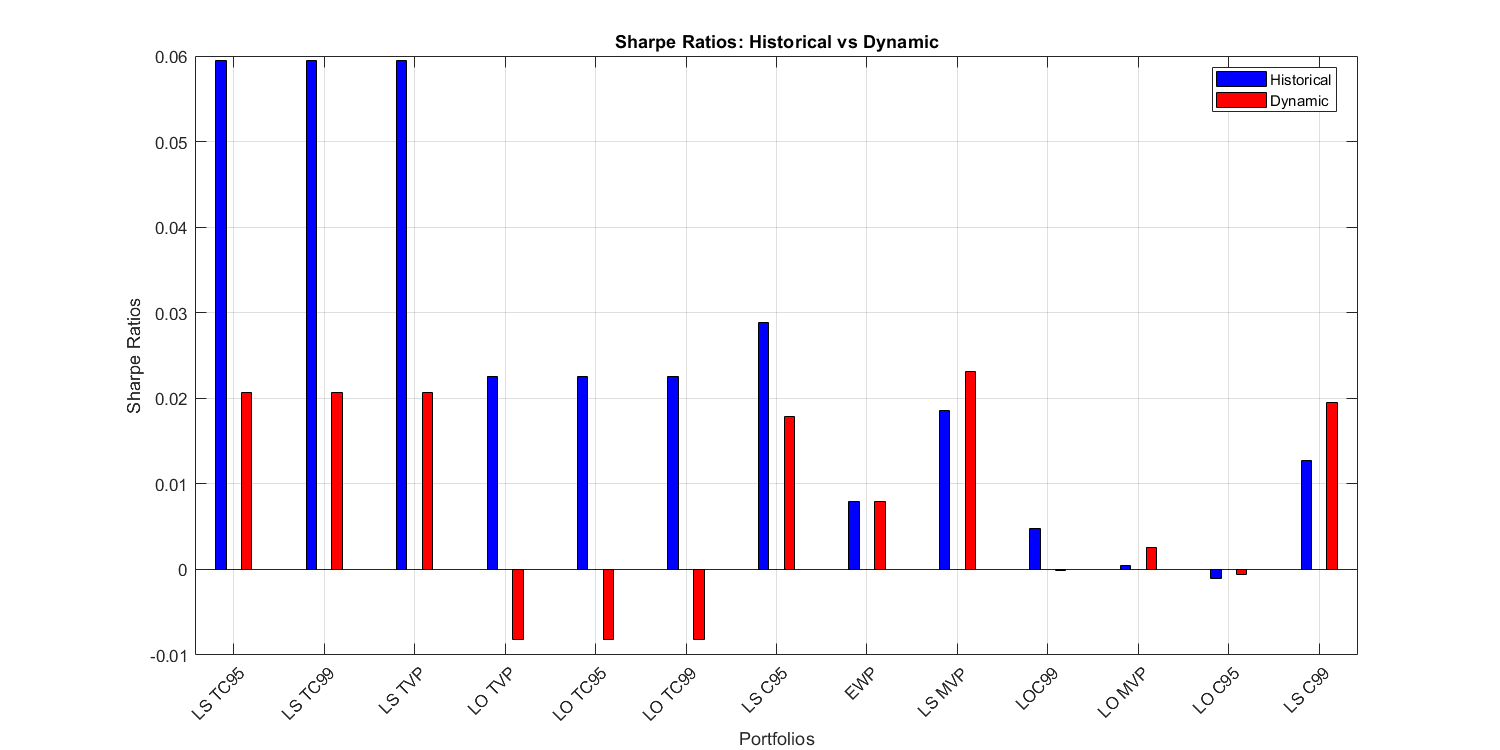}
\caption[The system.]{Historical vs. dynamic Sharpe ratios.}
\label{theSystemFig}
\end{figure}

\begin{figure}[H]
\captionsetup{justification=centering}
\centering
\includegraphics[width=1.0\textwidth]{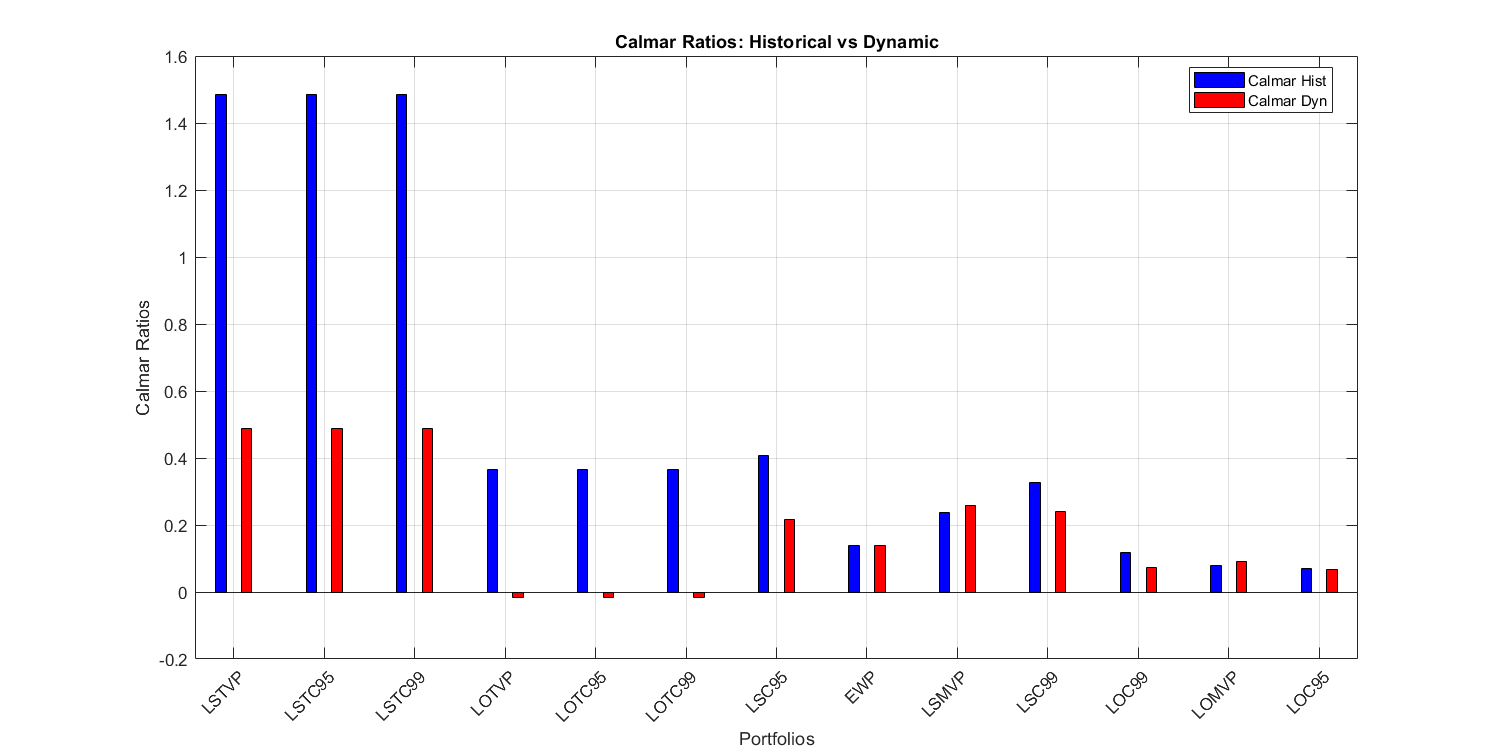}
\caption[The system.]{Historical vs. dynamic Calmar ratios.}
\label{theSystemFig}
\end{figure}

\begin{figure}[H]
\captionsetup{justification=centering}
\centering
\includegraphics[width=1.0\textwidth]{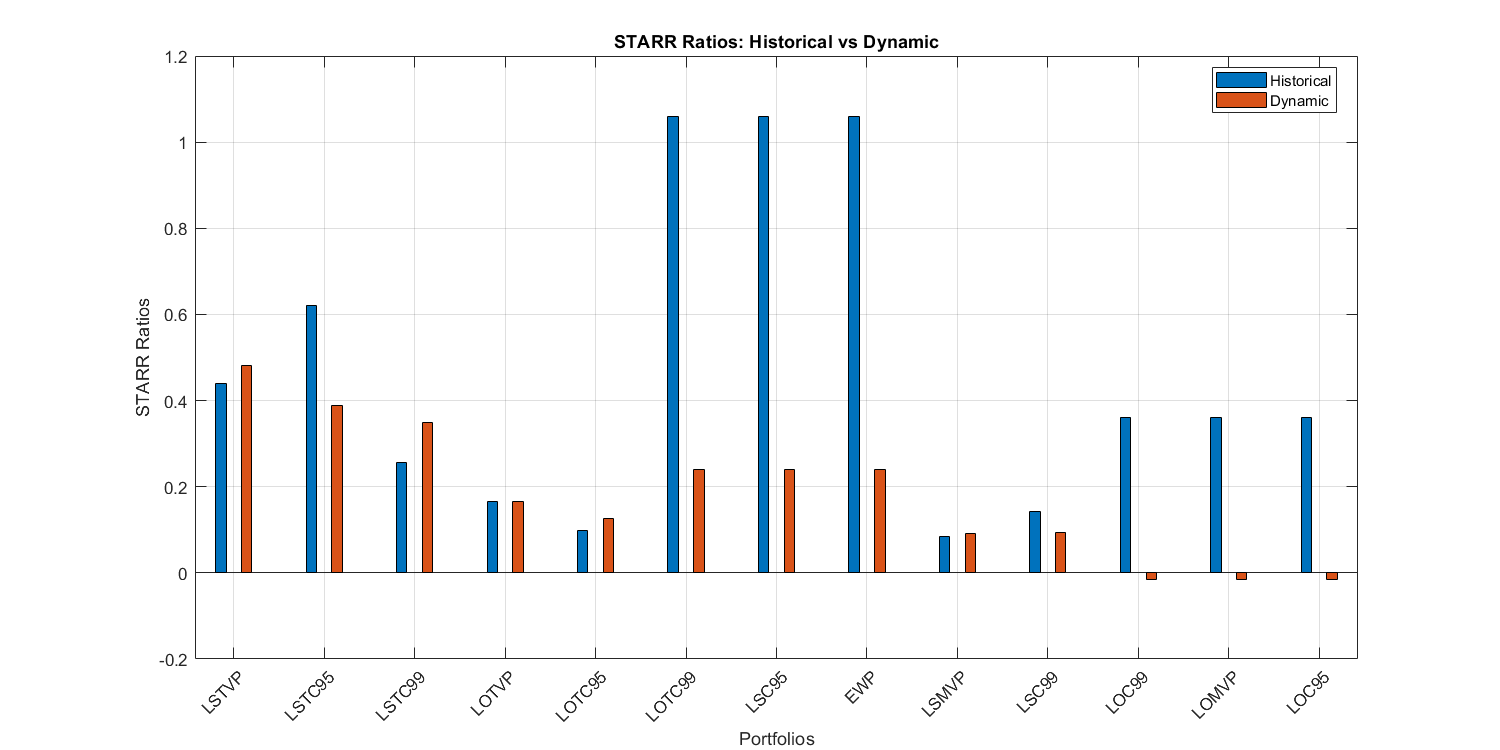}
\caption[The system.]{Historical vs. dynamic STARR ratios.}
\label{theSystemFig}
\end{figure}

\subsection{Historical ETFs vs. Historical DJIA Performance Ratios}

\begin{figure}[H]
\captionsetup{justification=centering}
\centering
\includegraphics[width=1.0\textwidth]{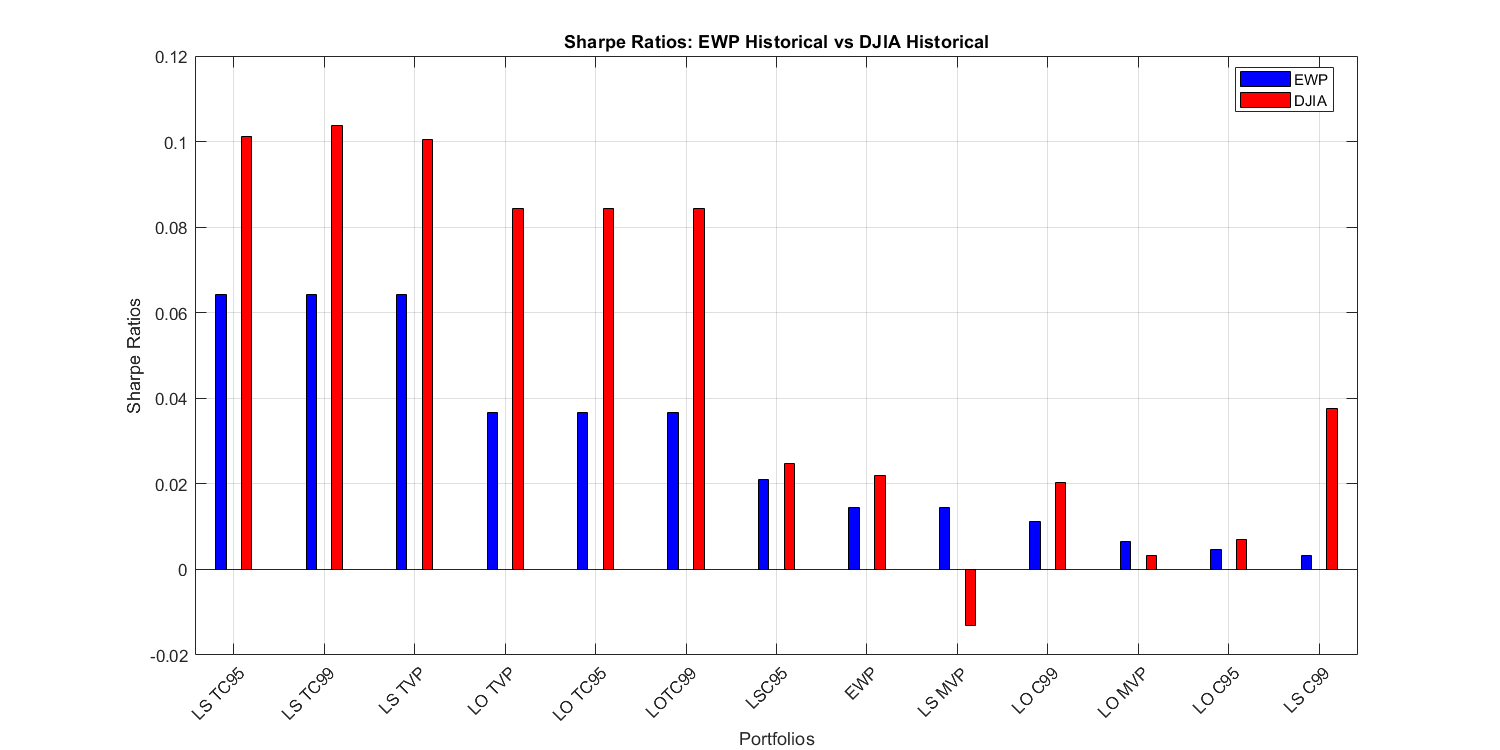}
\caption[The system.]{EWP vs. DJIA Sharpe ratios.}
\label{theSystemFig}
\end{figure}

\begin{figure}[H]
\captionsetup{justification=centering}
\centering
\includegraphics[width=1.0\textwidth]{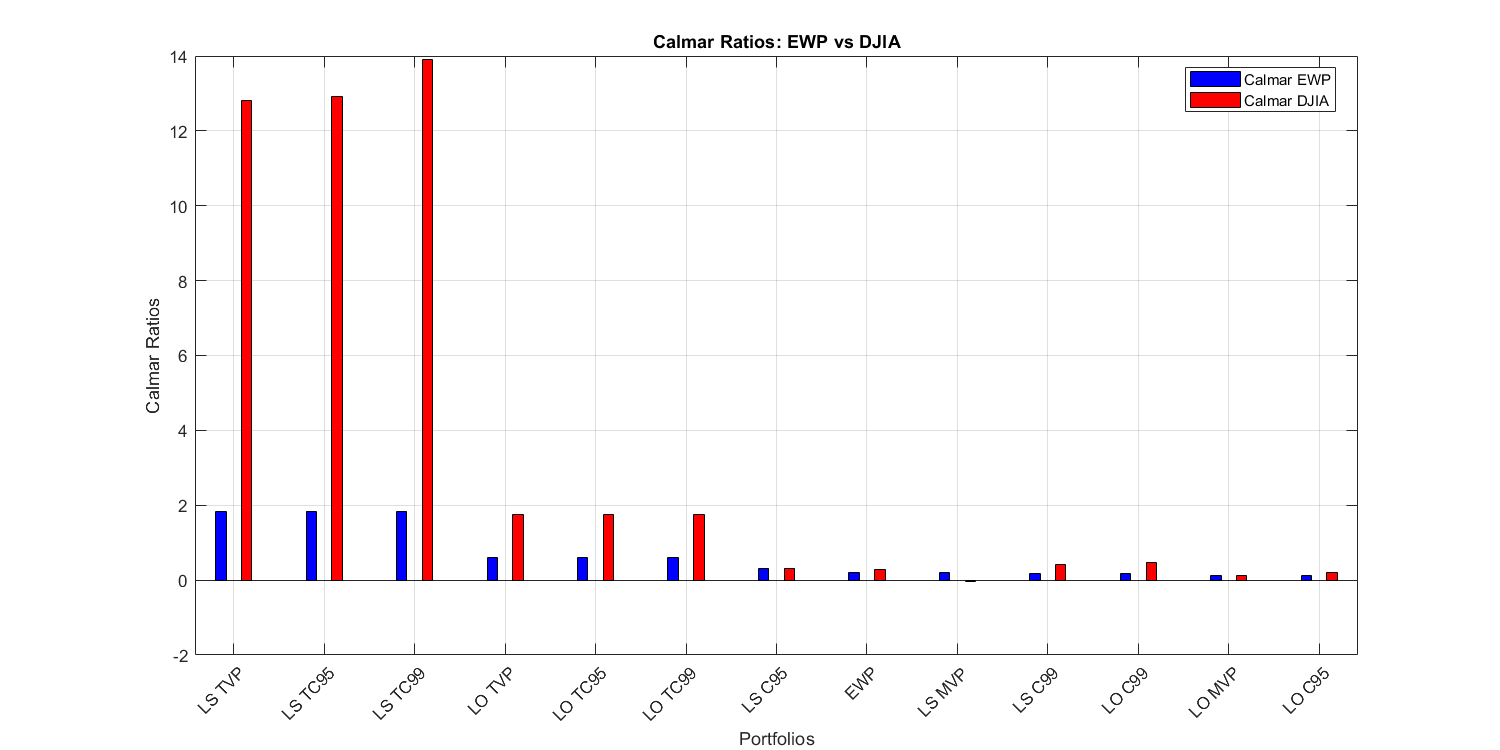}
\caption[The system.]{EWP vs. DJIA Calmar ratios.}
\label{theSystemFig}
\end{figure}

\begin{figure}[H]
\captionsetup{justification=centering}
\centering
\includegraphics[width=1.0\textwidth]{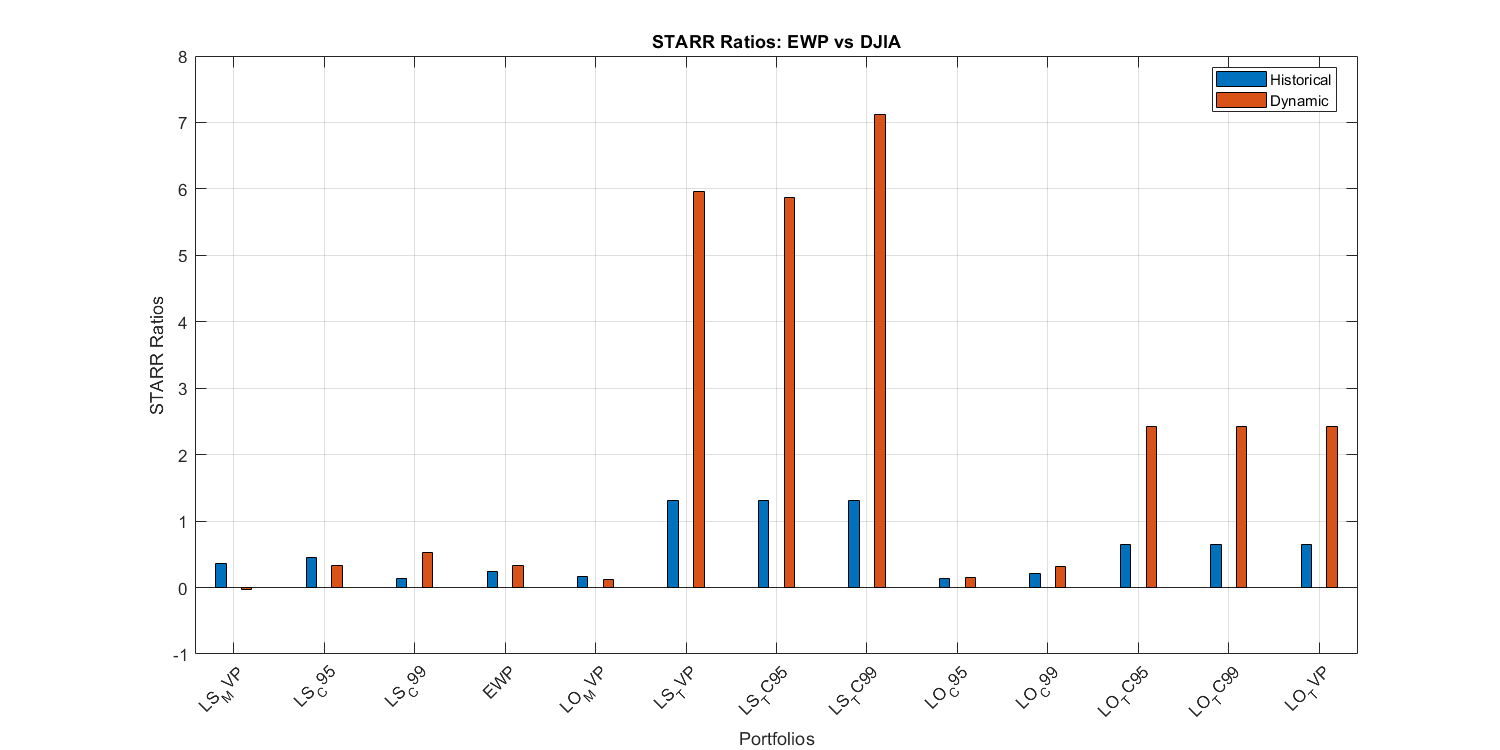}
\caption[The system.]{EWP vs. DJIA STARR ratios.}
\label{theSystemFig}
\end{figure}

\newpage
\bibliography{clean}

\begin{thebibliography}{12}
\providecommand{\natexlab}[1]{#1}
\providecommand{\url}[1]{\texttt{#1}}
\expandafter\ifx\csname urlstyle\endcsname\relax
  \providecommand{\doi}[1]{doi: #1}\else
  \providecommand{\doi}{doi: \begingroup \urlstyle{rm}\Url}\fi

\bibitem[Berger et~al.(2011)Berger, Pukthuanthong, and Yang]{Berger2011}
D.~Berger, K.~Pukthuanthong, and J.~J. Yang.
\newblock International diversification with frontier markets.
\newblock \emph{Journal of Financial Economics}, 101:\penalty0 227--242, 2011.
\newblock ISSN 0304405X.

\bibitem[Jilani(2024)]{Jilani2024}
J.~Jilani, Humza~Cotterill.
\newblock Frontier emerging markets stocks soar as investors cheer reforms.
\newblock \emph{Financial Times}, 2024.

\bibitem[Joyo and Lefen(2019)]{Joyo2019}
A.~S. Joyo and L.~Lefen.
\newblock Stock market integration of {P}akistan with its trading partners: A multivariate {DCC-GARCH} model approach.
\newblock \emph{Sustainability, (Switzerland)}, 11(2):\penalty0 303, 2019.
\newblock ISSN 20711050.

\bibitem[Lindquist et~al.(2021)Lindquist, Rachev, Hu, and Shirvani]{reit-portfolio}
W.~B. Lindquist, S.~T. Rachev, Y.~Hu, and A.~Shirvani.
\newblock \emph{Advanced REIT Portfolio Optimization: Innovative Tools for Risk Management}.
\newblock Springer, 2021.

\bibitem[Markowitz(1952)]{markowitz}
H.~Markowitz.
\newblock Portfolio selection.
\newblock \emph{The Journal of Finance}, 7(1):\penalty0 77--91, 1952.

\bibitem[Mensi et~al.(2017)Mensi, Shahzad, Hammoudeh, Zeitun, and Rehman]{Mensi2017}
W.~Mensi, S.~J.~H. Shahzad, S.~Hammoudeh, R.~Zeitun, and M.~U. Rehman.
\newblock Diversification potential of {A}sian frontier, {BRIC} emerging and major developed stock markets: A wavelet-based value at risk approach.
\newblock \emph{Emerging Markets Review}, 32:\penalty0 130--145, 2017.
\newblock ISSN 18736173.

\bibitem[Ngene et~al.(2018)Ngene, Post, and Mungai]{Ngene2018}
G.~Ngene, J.~A. Post, and A.~N. Mungai.
\newblock Volatility and shock interactions and risk management implications: Evidence from the {U.S.} and frontier markets.
\newblock \emph{Emerging Markets Review}, 37:\penalty0 66--79, 2018.
\newblock ISSN 18736173.

\bibitem[OECD(2019)]{oecd2019equity}
OECD.
\newblock Equity market review of {A}sia 2019.
\newblock OECD Capital Market Series, Paris, 2019.

\bibitem[PWC(2017)]{pwc2017}
PWC.
\newblock The long view.{H}ow will the global economic order change by 2050?, 2 2017.
\newblock PricewaterhouseCoopers.

\bibitem[Sharpe et~al.(1999)Sharpe, Alexander, and Bailey]{investments}
W.~F. Sharpe, G.~J. Alexander, and J.~V. Bailey.
\newblock \emph{Investments (6th ed.)}.
\newblock Prentice Hall, 1999.

\bibitem[Tsay(2005)]{tsay2005analysis}
R.~S. Tsay.
\newblock \emph{Analysis of Financial Time Series (2nd ed.)}.
\newblock John Wiley and Sons, 2005.

\bibitem[Woetzel et~al.(2018)Woetzel, Madgavkar, Seong, Manyika, and Sneader]{Woetzel2018}
J.~Woetzel, A.~Madgavkar, J.~Seong, J.~Manyika, and K.~Sneader.
\newblock Outperformers: High-growth emerging economies and the companies that propel them.
\newblock McKinsey Global Institute, 2018.

\end{thebibliography}
\newpage

\end{document}